# A reduced-order modeling framework for simulating signatures of faults in a bladed disk

Divya Shyam Singh[1,a], Atul Agrawal[1,a,b], D. Roy Mahapatra[a]

[a]Department of Aerospace Engineering, Indian Institute of Science, Bangalore 560012, India

[b]Department of Mechanical Engineering, Technical University of Munich, Boltzmannstrasse 15, 85747 Munich, Germany

**Abstract**

This paper reports a reduced-order modeling framework of bladed disks on a rotating shaft to simulate the vibration signature of faults in different components aiming toward simulated data-driven machine learning. We have employed lumped and one-dimensional analytical models of the subcomponents for better insight into the complex dynamic response. The framework addresses some of the challenges encountered in analyzing and optimizing fault detection and identification schemes for health monitoring of aero-engines and other rotating machinery. We model the bladed disks and shafts by combining lumped elements and one-dimensional finite elements, leading to a coupled system. The simulation results are in good agreement with previously published data. We model and analyze the cracks in a blade with their effective reduced stiffness approximation. Different types of faults are modeled, including cracks in the blades of a single and a two-stage bladed disks, Fan Blade Off (FBO), and Foreign Object Damage (FOD). We have applied aero-engine operational load conditions to simulate realistic scenarios of online health monitoring. The proposed reduced-order simulation framework will have applications in probabilistic signal modeling, machine learning toward fault signature identification, and parameter estimation with measured vibration signals.

## 1. Introduction

For decades, health monitoring system to detect faults in rotors has remained a significant area of interest, particularly to the aero-engine industry and turbine-based power plant operators. Structural Health Monitoring (SHM) system design typically deals with performance-related challenges involving sensors, interfaces, and software algorithms to efficiently function under harsh environmental conditions and detect faults from noisy and complex signals. Verification and validation of the overall system performance through modeling signals have great promises. Simulations can help reduce the design and development time and minimize costly experiments with actual hardware. Health monitoring may also involve many non-destructive inspections in correlating online or condition-based information and helping maintenance decisions. A typical monitoring task requires reporting faults and their possible type. A reliability-based estimate of the inspection outcome based on the specific schemes of fault identification (such as the pattern of occurrence, signal parameter correlation etc.) may be developed. Once a fault is reported and identified to have been caused by structural damage, the objective is to locate the damage(s), identify its nature and severity in terms of some representative parameters or indicators. However, in the case of complex geometries and large structures, it can be costly as well as impractical. This limitation gives rise to a need for reliable algorithms to indirectly monitor the health of the components from performance or functional response. An example of such a simplification of fault detection and identification process is in bladed disk systems where faults in the blades and blade roots are to be identified from online monitoring of vibration data from bearings, shaft, and the casing of turbine systems etc. Madhavan [1] used the vibration data from the rotor's blade using sensors mounted on the turbine casing to detect damage. Ou [2] used the force sensors, and accelerometers mounted on a wind turbine blade to monitor the wind turbine under various conditions in an experimental setup. Thus, patterns in the signals may comparatively indicate the state of vibration of the shaft and the neighbouring bladed disks (in the case of a multi-stage bladed disk).

---

[1] Contributed equally

For the health monitoring of an aero-engine, multiple different sensors are used to measure various physical parameters. The vibration parameters of the shaft and the blades are one of the most critical parameters to be measured and analyzed for health monitoring of a bladed disk. Efforts to monitor aero-engines began as early as the 1940s [3], where magnets were used mounted on the blade end to measure vibrations which were known as frequency modulated grid. An apparatus was designed by Luongo [4] for the detection of vibration in the blade, which used multiple sensors placed evenly in one row. With this, a particular blade could be monitored. However, these methods were costly, and when the number of blades increases, such as in the case of a multi-stage turbine, these sensors become a less attractive option. Late in the 1980s, other methods for monitoring the blade vibration using optical or capacitance-based sensors were adopted, providing the vibration data for all the blades in a stage with just a single sensor. A method of finding out blade vibration using optical sensors was reported by Zielinski and Ziller [5]. Mollmann [4] used the rotation of gear connected to the shaft for the turbine shaft vibration response. A sensor produced a pulse train because of the gear teeth, and it was used to calculate the RPM and the vibration. These methods of sensing the vibration of the turbine blade components aim to provide the machine's live operating conditions. A fault in the system can be detected based on these signals using an appropriate fault detection scheme.

The fault detection methodology can be broadly divided into the signal model-based, data-driven, and physical model-based approaches. The signal model-based approach uses the change in the measured signal from the system for fault diagnosis. Once the change in the measured signal is detected, it may be followed by a feature extraction process. Based on the extracted features, a decision can then be made based on prior knowledge regarding such features. The simplest form of signal-based fault detection is checking the state of the system based on the intensity or peaks of the signal [6]. Wang et al. [7] carried out the fault location and severity prediction from the frequency response function (FRF). The accuracy of their method was checked on numerical as well as experimental models of a three-bay frame structure. In ref. [8], an array of eleven Kalman filters was used for fault isolation for actuators and sensors. Each of the filters was designed to detect a specific sensor fault. It was demonstrated that the algorithm could detect the sensor or actuator faults in an aero-engine even in the presence of system faults. Hou et al. [9] carried out SHM and damage detection on a simulated structural model using a wavelet-based approach. They demonstrated the applicability of a simple spring-mass model and vibration data of buildings during the event of an earthquake. However, detecting small-sized faults from signal-based methods is a challenging task as feature extraction from the signal becomes problematic in the presence of measurement noise.

A data-driven method for fault detection has increasingly become popular in recent years. Numerous data-driven techniques have been applied for fault detection and health monitoring. Wu and Liu [10], [11] used wavelet transformation coupled with artificial neural networks to detect faults such as misfiring in cylinders, air leakage, and sensor faults in an IC engine. A hybrid network of Support Vector Machine (SVM) and Artificial Neural Network (ANN) was used for fault diagnosis by Seo et al. [12]. The advantage of using this hybrid network was the reduction in training data required and the time necessary to train the network. SVM was used to detect the fault's location, and the ANN estimated the magnitude. This methodology was used for fault detection in a gas turbine engine for an unmanned air vehicle. Wu et al. [13] proposed a novel fault prognosis (remaining useful life prediction) scheme using a recurrent neural network that was applied to an aircraft turbo-fan engine. However, the applicability of these techniques in the real-time monitoring of systems is still an area of research because of the computational resource and time required.

The physical model-driven approach relies on the level of approximation of the physics and mathematical models used. It involves monitoring system response and comparing it with that of the model form. The change in the difference can then be traced back to a probable fault in terms of the system parameters, system input or sensor measurements, or combinations. Several algorithms can be applied to model-based fault detection and identification, some of which are briefly discussed here.

Observer-based fault detection is a model-based fault detection method in which a state-space model is constructed, and the fault detection is carried out through residual or error calculation. The residual value depends on the process fault as well as external disturbances. Ideally, the residual signal should be sensitive to faults and insensitive to external disturbances. One of the earliest observer-based fault detection schemes for aircraft sensor fault analysis was introduced by Patton, Willcox and Winter [14] demonstrated the feasibility in real-time using observation error subspace. Dai et al. [15] developed a computationally efficient fault detection observer, making it suitable for online health monitoring. They demonstrated the improvement in the actuator fault detection of a gas turbine engine through the improved observer. Kulesza et al. [16] used the unknown input observer-based fault detection to detect shallow cracks in a shaft. A specially designed linear matrix inequality-based technique was used for $H_-/H_\infty$ optimization which helped to arrive at the optimal solution faster. Their method was validated against numerical simulation as well as experimental results. A fault detection technique using a high-gain observer was proposed by Gao et al. [17], and it was applied on a gas turbine engine. Observer-based fault detection methods are more straightforward than model-based fault detection, which uses a physics-based model of the system. Physics-based models require a parametric representation of the fault to be incorporated into the model.

Majority of the studies involving an analytical model of the bladed disc deal with the problem of mistuning in the blade. Mistuning in blades occurs when one or more blades are unidentical to the other, potentially leading to instabilities and unusually high vibrations. There have been several efforts to model the turbine blade system by modeling individual blades as Euler-Bernoulli beams connected to the rigid shaft. The mistuning is introduced as a change in the stiffness of one or some of the blades. Dye and Henry reported one of the earliest studies on blade mistuning [18]. Using a lumped model, they examined the coupling among the blades of a gas turbine compressor blades. One blade can have higher stresses if the natural frequency is slightly different from the other blades, giving rise to resonance. However, Ewins [19] found that approximating the blade with a simple cantilever beam overlooks many natural frequencies. He discovered that blade detuning could give rise to irregular vibration in a turbine blade system. Using a numerical model of the bladed disk with a cracked blade near the root, Huang [20] investigated the effect of mode localization. Efforts have also been made by assuming the disk as a circular plate element and beam elements to represent the blades [21] to study the vibration and flutter response of the blade. Using a classical fracture mechanics-based approach, Fang et al. [22] studied the effect of a crack on the turbine blade system response. Free and forced response of the damaged bladed disk was calculated, and it was found that even a tiny crack was capable of giving rise to the mode localization effect. Lim et al. [23] developed a generic framework of the reduced-order model of the mistuned bladed disk, which can be applied to small and large mistuning problems. They divided their model into a tuned model and a virtual mistuned blade, where the mistuned component was defined as the difference in the mass and stiffness matrices of the mistuned and tuned blades. The impact of crack depth, twist angle, and engine spend on the localization effect due to crack was investigated by Kuang and Huang [24]. Another common approach to modeling the mistuning in the bladed disk is to use a detailed finite element model that provides much greater accuracy but it is computationally prohibitive. Saito et al. [25] developed a nonlinear crack model to account for the effect of crack opening and closing. The response was calculated based on a finite element model. The forced response of the blade was calculated, treating the crack blade as a cantilever. Xu [26] studied the effect of a breathing crack on a mistuned rotor blade system numerically based on the change in natural frequency, vibration amplitude etc.. Hou [27] developed an analytical lumped parameter model where the crack was introduced as a local reduction in the stiffness. The effect of mistuning induced by the crack on the natural frequency was studied. Most of the works described above either use a reduced-order finite element model or a detailed finite element model to represent the bladed disks, followed by model reduction techniques to make the problem computationally tractable. The damage effect on the blade was studied, assuming the blades were to be attached to a rigid shaft, thus treating the blades as cantilever beams.

Other types of defects common in a turbine blade system are foreign object damage (FOD) and Fan blade out (FBO). FOD event occurs when small-sized foreign objects such as birds collide with the turbine blade. The majority of FOD cases are not fatal; however, sometimes, they can induce instability in the system or create cracks [27], which might require corrective actions. Turso et al. [28] extracted the features of a FOD event on a turbo-fan model. For test data generation for a FOD event, a reduced-order turbo-fan model was developed through a finite element code called DyRoBes. They were able to precisely detect the occurrence of FOD events in time using a wavelet-based feature extraction method.

FBO events pose a serious threat to the safety of an aircraft. The sudden force created because of the imbalance generates extreme loads in the shafts and the flown off the blade, which can lead to further damage to the aircraft [29]. Yu et al. [30] proposed a dynamic modeling method of an aero-engine in FBO events. They studied the system response for an FBO event with blade rubbing with rotor casing considered. A finite element software to model the system was used. The system properties from the finite element model were extracted, and subsequently, Newmark's time integration scheme was used to calculate the system response.

Most of the studies reported in the literature attempt to model the turbine blade system in a reduced framework to address the effect mistuning has on the forced vibration of blades. On the other hand, the model of the turbine blade system to deal with damages such as FOD and FBO cannot capture the system response in the presence of cracks in the blade since the blades are modeled as lumped mass at the disk circumference.

The contribution of the present paper is a modeling framework capable of accommodating the impact of various types of faults on the response of the turbine shaft using a reduced-order three-dimensional finite element model. We have studied the effects of three different damage cases on the system response, which are (1) cracks in the blades, (2) FBO, and (3) FOD. The cracks are modeled using a concept proposed by Christides and Barr [31], which utilizes a simple approximation for a local reduction in stiffness. For simulating the fan blade off event, the length of one of the blades was reduced abruptly, and the foreign object damage was simulated by applying a sudden impact force at the blade tip. The shaft response for all the damage cases was simulated and analyzed. These details in the paper have been organized in the following manner. Section 2 elucidates the modeling scheme applied for the turbine blade system and the cracks, followed by the simulation scheme used. Section 3 is divided into three sections, each of which deals with one case of damage, as explained above. The paper is concluded in Section 4, and some grounds for future works are discussed.

## 2. Modelling Framework

### 2.1. Reduced-order modeling of bladed disk

The shaft is modelled as a one-dimensional element with a hollow cylindrical cross-section. The disk is modelled as lumped mass. The blades are modeled with beam elements, forming a coupled system of equations involving external aerodynamic load (Figure 1(a)). The disk is modeled as a lumped mass attached to the end of the shaft. The blade roots are also assumed to be connected at the shaft end. The global coordinate system has its origin at the beginning of the shaft, whereas each of the blades has its local coordinate system origin situated at the blade root. Appropriate transformation is required for the assembly of the blade element equations into the global matrix-vector system equation. A turbine blade system with one element per blade and shaft is shown in Figure 1(b). Superscript S indicates shaft, and superscript B indicates blade. The displacement DOFs of the shaft vibration at an element node-1 and 2 in the global reference frame are expressed as $\boldsymbol{u}_i^S = \{u_X^S \ u_Y^S \ u_Z^S \ \theta_X^S \ \theta_Y^S \ u_Z^S\}^T, i = 1,2$. The displacement DOFs at an element node of the $j^{th}$ blade is expressed as $\boldsymbol{u}_{i,j}^B = \{u_X^B \ u_Y^B \ u_Z^B \ \theta_X^B \ \theta_Y^B \ u_Z^B\}^T, i = 1,2$. $u$ indicates translational displacement, and $\theta$ indicates rotation angle. Corresponding element force vectors $\boldsymbol{f}^S$ and $\boldsymbol{f}_j^B$ are approximated depending on various sources of loads which are discussed later.

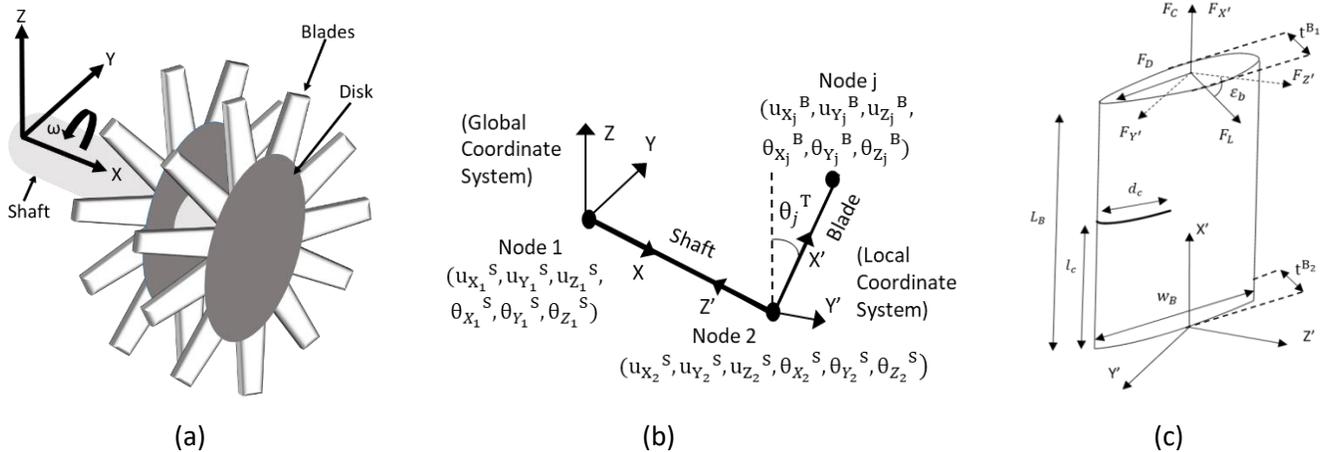

(a)          (b)          (c)

Figure 1 (a) Multi-Stage Turbine blade system (b) Finite element model of the turbine blade system showing j<sup>th</sup> blade and the corresponding nodes (c) Isometric view of the turbine blade showing the aerodynamic forces, blade and crack dimensions

The global system of the finite element equations can be written as

$$M\ddot{u}(t) + C\dot{u}(t) + Ku(t) = f(t) \tag{1}$$

which is obtained as finite element coordinate transformations and assembly of elements of shaft and blades as (see Fig. 2)

$$\begin{bmatrix} M^S & \cdots & \cdots \\ \vdots & \ddots & \vdots \\ \cdots & \cdots & M_j^B \end{bmatrix} \begin{Bmatrix} \ddot{u}^S \\ \vdots \\ \ddot{u}_j^B \end{Bmatrix} + \begin{bmatrix} C^S & \cdots & \cdots \\ \vdots & \ddots & \vdots \\ \cdots & \cdots & C_j^B \end{bmatrix} \begin{Bmatrix} \dot{u}^S \\ \vdots \\ \dot{u}_j^B \end{Bmatrix} + \begin{bmatrix} K^S & \cdots & \cdots \\ \vdots & \ddots & \vdots \\ \cdots & \cdots & K_j^B \end{bmatrix} \begin{Bmatrix} u^S \\ \vdots \\ u_j^B \end{Bmatrix} = \begin{Bmatrix} f^S \\ \vdots \\ f_j^B \end{Bmatrix} \tag{2}$$

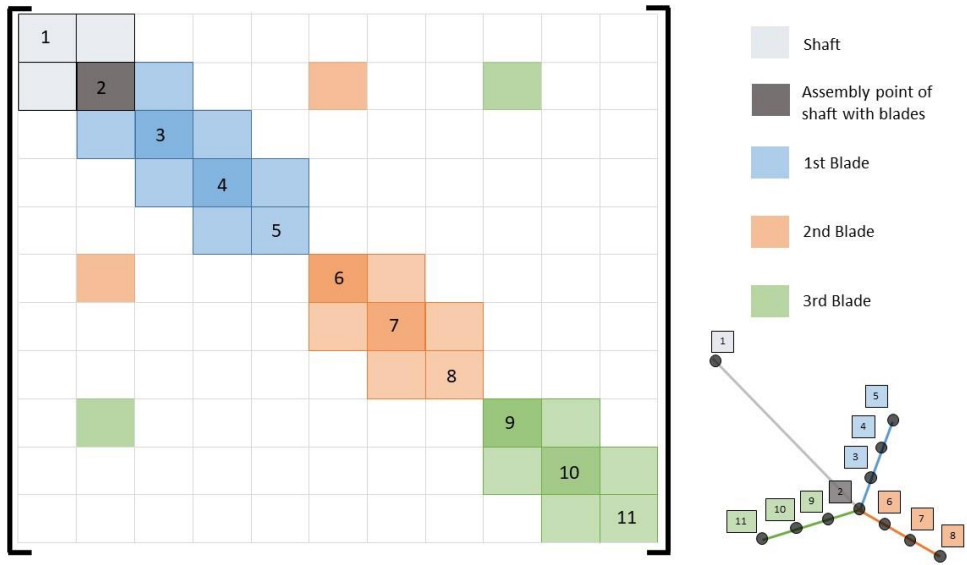

Figure 2 Assembly pattern of the finite element matrices for a system with 3 blades, 3 elements per blade and 1 element in shaft.

The global matrices are essentially constructed from the assembly of the matrices for the shaft and the blade. The matrices making up the full system matrices are now discussed. The shaft's mass and stiffness matrices and displacement vector in global Y and Z directions are first constructed. Since the local and global coordinate system for the shaft is the same, the mass, stiffness matrices, and displacement vector in the global coordinate system can be written as

$$M_Y^S = \frac{\rho A^S L^S}{420} \begin{bmatrix} 156 & 22L^S & 54 & -13L^S \\ 22L^S & 4L^{S^2} & 13L^S & -3L^{S^2} \\ 54 & 13L^S & 156 & -22L \\ -13L^S & -3L^{S^2} & -22L^S & 4L^{S^2} \end{bmatrix}, K_Y^S = \frac{EI_Y^S}{L_S^3} \begin{bmatrix} 12 & 6L^S & -12 & 6L^S \\ 6L^S & 4L^{S^2} & -6L^S & 2L^{S^2} \\ -12 & -6L^S & 12 & -6L^S \\ 6L^S & 2L^{S^2} & -6L^S & 4L^{S^2} \end{bmatrix}, u_Y^S = \begin{pmatrix} u_{Y_1}^S \\ \theta_{Z_1}^S \\ u_{Y_2}^S \\ \theta_{Z_2}^S \end{pmatrix} \quad (3)$$

$$M_Z^S = \frac{\rho A^S L^S}{420} \begin{bmatrix} 156 & 22L^S & 54 & -13L^S \\ 22L^S & 4L^{S^2} & 13L^S & -3L^{S^2} \\ 54 & 13L^S & 156 & -22L \\ -13L^S & -3L^{S^2} & -22L^S & 4L^{S^2} \end{bmatrix}, K_Z^S = \frac{EI_Z^S}{L_S^3} \begin{bmatrix} 12 & 6L^S & -12 & 6L^S \\ 6L^S & 4L^{S^2} & -6L^S & 2L^{S^2} \\ -12 & -6L^S & 12 & -6L^S \\ 6L^S & 2L^{S^2} & -6L^S & 4L^{S^2} \end{bmatrix}, u_Z^S = \begin{pmatrix} u_{Z_1}^S \\ \theta_{Y_1}^S \\ u_{Z_2}^S \\ \theta_{Y_2}^S \end{pmatrix} \quad (4)$$

The mass and stiffness matrices and displacement vector for longitudinal and torsion in the global X direction can be written as

$$M_X^S = \begin{bmatrix} \frac{\rho A^S L^S}{3} & 0 & \frac{\rho A^S L^S}{6} & 0 \\ 0 & \frac{\rho J_X^S L^S}{3} & 0 & \frac{\rho J_X^S L^S}{6} \\ \frac{\rho A^S L^S}{6} & 0 & \frac{\rho A^S L^S}{3} & 0 \\ 0 & \frac{\rho J_X^S L^S}{6} & 0 & \frac{\rho J_X^S L^S}{3} + J_X^D \end{bmatrix}, K_X^S = \begin{bmatrix} \frac{EA^S}{L^S} & 0 & -\frac{EA^S}{L^S} & 0 \\ 0 & \frac{GJ_X^S}{L^S} & 0 & -\frac{GJ_X^S}{L^S} \\ -\frac{EA^S}{L^S} & 0 & \frac{EA^S}{L^S} & 0 \\ 0 & -\frac{GJ_X^S}{L^S} & 0 & \frac{GJ_X^S}{L^S} \end{bmatrix}, u_X^S = \begin{pmatrix} u_{X_1}^S \\ \theta_{X_1}^S \\ u_{X_2}^S \\ \theta_{X_2}^S \end{pmatrix} \quad (5)$$

The area of the shaft, the mass, and the polar moment of inertia of the disk lumped at the shaft end are given by

$$A^S = \pi \left( \frac{d^{S_o 2}}{4} - \frac{d^{S_i 2}}{4} \right), M^D = \rho^D t^D \pi \left( \frac{d^{D^2}}{4} - \frac{d^{S_o 2}}{4} \right), J_X^D = \frac{\pi}{2} \left( \frac{d^{D^4}}{16} - \frac{d^{S_o 4}}{16} \right) \quad (6)$$

where $d^{S_o}$ is the outer diameter of the shaft, $d^{S_i}$ is the inner diameter of the shaft, $d^D$ is the diameter of the disk, $\rho^D$ is the density of the disk and $t^D$ is the thickness of the disk. The area moment of inertia and polar moment of inertia of the shaft is calculated as

$$I_Y^S = I_Z^S = \frac{\pi}{4} \left( \frac{d^{S_o 4}}{16} - \frac{d^{S_i 4}}{16} \right), J_X^S = \frac{\pi}{2} \left( \frac{d^{S_o 4}}{16} - \frac{d^{S_i 4}}{16} \right) \quad (7)$$

The above matrices are assembled into a single matrix to form the complete matrix for the shaft for mass and stiffness and the final unknown degree of freedom vector. The details of the entire system matrices are given in Appendix-A. For the calculation of the damping matrix, a two-step formulation given by Chowdhury and Dasgupta [32] was followed. Similarly, the mass and stiffness matrices and displacement vector for bending of the blade in local Y' and Z' directions can be written as

$$\boldsymbol{M}_Y^{B'} = \frac{\rho A^B L_e^B}{420} \begin{bmatrix} 156 & 22L_e^B & 54 & -13L_e^B \\ 22L_e^B & 4L_e^{B^2} & 13L_e^B & -3L_e^{B^2} \\ 54 & 13L_e^B & 156 & -22L_e^B \\ -13L_e^B & -3L_e^{B^2} & -22L_e^B & 4L_e^{B^2} \end{bmatrix}, \quad \boldsymbol{K}_Y^{B'} = \frac{EI_{Y'}^B}{L_e^{B^3}} \begin{bmatrix} 12 & 6L_e^B & -12 & 6L_e^B \\ 6L_e^B & 4L_e^{B^2} & -6L_e^B & 2L_e^{B^2} \\ -12 & -6L_e^B & 12 & -6L_e^B \\ 6L_e^B & 2L_e^{B^2} & -6L_e^B & 4L_e^{B^2} \end{bmatrix}, \quad \boldsymbol{u}_Y^{B'} =$$

$$\begin{pmatrix} u_{Y_1}^B & \theta_{Z_1}^B & u_{Y_1}^B & \theta_{Z_2}^B \end{pmatrix}^T \tag{8}$$

$$\boldsymbol{M}_Z^{B'} = \frac{\rho A^B L_e^B}{420} \begin{bmatrix} 156 & 22L_e^B & 54 & -13L_e^B \\ 22L_e^B & 4L_e^{B^2} & 13L_e^B & -3L_e^{B^2} \\ 54 & 13L_e^B & 156 & -22L_e^B \\ -13L_e^B & -3L_e^{B^2} & -22L_e^B & 4L_e^{B^2} \end{bmatrix}, \quad \boldsymbol{K}_Z^{B'} = \frac{EI_{Z'}^B}{L_e^{B^3}} \begin{bmatrix} 12 & 6L_e^B & -12 & 6L_e^B \\ 6L_e^B & 4L_e^{B^2} & -6L_e^B & 2L_e^{B^2} \\ -12 & -6L_e^B & 12 & -6L_e^B \\ 6L_e^B & 2L_e^{B^2} & -6L_e^B & 4L_e^{B^2} \end{bmatrix}, \quad \boldsymbol{u}_Z^{B'} =$$

$$\begin{pmatrix} u_{Z_1}^B & \theta_{Y_1}^B & u_{Z_1}^B & \theta_{Y_2}^B \end{pmatrix}^T \tag{9}$$

The mass and stiffness matrices and displacement vectors for extension and torsion in the local X direction can be written as

$$\boldsymbol{M}_X^{B'} = \begin{bmatrix} \frac{\rho A^B L_e^B}{3} & 0 & \frac{\rho A^B L_e^B}{6} & 0 \\ 0 & \frac{\rho J_X^B L_e^B}{3} & 0 & \frac{\rho J_X^B L_e^B}{6} \\ \frac{\rho A^B L_e^B}{6} & 0 & \frac{\rho A^B L_e^B}{3} & 0 \\ 0 & \frac{\rho J_X^B L_e^B}{6} & 0 & \frac{\rho J_X^B L_e^B}{3} \end{bmatrix}, \boldsymbol{K}_X^{B'} = \begin{bmatrix} \frac{EA^B}{L_e^B} & 0 & -\frac{EA^B}{L_e^B} & 0 \\ 0 & \frac{GJ_X^B}{L_e^B} & 0 & -\frac{GJ_X^B}{L_e^B} \\ -\frac{EA^B}{L_e^B} & 0 & \frac{EA^B}{L^B} & 0 \\ 0 & -\frac{GJ_X^B}{L_e^B} & 0 & \frac{GJ_X^B}{L_e^B} \end{bmatrix}, \boldsymbol{u}_X^{B'} = \begin{pmatrix} u_{X_1}^B \\ \theta_{X_1}^B \\ u_{X_1}^B \\ \theta_{X_2}^B \end{pmatrix} \tag{10}$$

The area, area moment of inertia, and polar moment of inertia for the blade with an approximately trapezoidal cross-section assumed are

$$A^B = \frac{1}{2}(t^{B_1} + t^{B_2})w^B, \quad I_{Y'}^B = \frac{(t^{B_1^2} + 4t^{B_2}t^{B_1} + t^{B_2^2})w^{B^3}}{36(t^{B_1} + t^{B_2})}, \quad I_{Z'}^B = \frac{w^B(t^{B_1} + t^{B_2})(t^{B_1^2} + t^{B_2^2})}{48}, \quad J_{X'}^B = I_{Y'}^B + I_{Z'}^B \tag{11}$$

where $t^{B_1}$ and $t^{B_2}$ are the lengths of the parallel sides of the trapezoid and $w^B$ is the width of the blade. The complete matrices have been listed in Appendix-A

### 2.1.1. Damage Modelling

Blades develop surface cracks on exterior surfaces in thin solid blades and in hollow interior surfaces, both of which are assumed to be equivalent cracks in the chord-wise section of the blade due to rotational and aerodynamic loads. Here, the cracks are considered to reduce the local stiffness of the blade cross-section against axial, bending, and torsional loads. A detailed model of local geometry and crack geometry could be quite complicated. However, it can be approximated as a local flexibility change near the cracked area in terms of reduction of the area moment, cross-sectional area, and polar moment. We employed an exponential solution to a decrease in the stiffness near the cracked zone suggested by Christides and Barr [31]. The model takes into consideration both crack location and crack depth quite accurately. The effective flexural rigidity for an approximated blade cross-section as a trapezoidal cross-section of a beam with a crack for bending in Z′ and Y′ directions is given by

$$EI_{Z'}^B(x) = EI_{Z'}^B \left[1 + Ce^{\left(\frac{-2\gamma_0|x-l_c|}{w^B}\right)}\right]^{-1}, \quad C = \frac{I_{Z'}^B - I_{Z'}^{B_C}}{I_{Z'}^{B_C}}, \quad I_{Z'}^B = \frac{1}{48}w^B(t^{B_1} + t^{B_2})(t^{B_1^2} + t^{B_2^2}),$$

$$I_{Z'}^{B_C} = \frac{1}{48}(w^B - d_c)(t^{B_1} + t^{B_2})(t^{B_1^2} + t^{B_2^2}) \tag{12}$$

$$EI_{Y'}^B(x) = EI_{Y'}^B \left[1 + Ce^{\left(\frac{-2\gamma_0|x-l_c|}{w^B}\right)}\right]^{-1}, \quad C = \frac{I_{Y'}^B - I_{Y'}^{B_C}}{I_{Y'}^{B_C}}, \quad I_{Y'}^B = \frac{(t^{B_1^2} + 4t^{B_2}t^{B_1} + t^{B_2^2})w^{B^3}}{36(t^{B_1} + t^{B_2})},$$

$$I_{Y'}^{B_C} = \frac{(t^{B_1^2} + 4t^{B_2}t^{B_1} + t^{B_2^2})(w^B - d_c)^3}{36(t^{B_1} + t^{B_2})} \tag{13}$$

where $l_c$ is the crack location, $\gamma_0$ is an experimentally determined parameter which is taken as 0.667, $d_c$ is the crack depth, $w^B$ is the width of the blade and $t^B$ is the thickness of the blade. The functions $I_{Z'}^B(x), I_{Y'}^B(x)$ are defined in the domain $\in [0, L^B]$ where $L^B$ is the total length of the blade. The shape functions are defined in the domain $\in [0, L_e^B]$ where $L_e^B$ is the length of an element of the blade, the coordinate system of the functions $I_{Z'}^B(x), I_{Y'}^B(x)$ have to be shifted for each element in the following manner before integration

$$I_{Z'}^B(x) = I_{Z'}^B \left[1 + Ce^{\left(\frac{-2\gamma_0|x+(i-1)L_e^B - l_c|}{w^B}\right)}\right]^{-1}, \quad I_{Y'}^B(x) = I_{Y'}^B \left[1 + Ce^{\left(\frac{-2\gamma_0|x+(i-1)L_e^B - l_c|}{w^B}\right)}\right]^{-1} \tag{14}$$

where $i \in [1,2,3 \dots n]$ where n is the total number of elements in the blade. Similarly, the longitudinal stiffness of the blade is approximated as

$$EA(x) = EA^B \left[1 + Ce^{\left(\frac{-2\gamma_0|x-l_c|}{d}\right)}\right]^{-1}, \quad C = \frac{A^B - A^{B_C}}{A^{B_C}}, \quad A^B = \frac{1}{2}(t^{B_1} + t^{B_2})w^B, \quad A^{B_C} = \frac{1}{2}(t^{B_1} + t^{B_2})(w^B - d_c) \tag{15}$$

Similarly, since the function $A^B(x)$ is defined in the domain $\in [0, L^B]$ where $L^B$ is the total length of the blade and the shape functions are defined in the domain $\in [0, L_e^B]$ where $L_e^B$ is the length of an element of the blade, the coordinate system of the function $A^B(x)$ has to be shifted for each element in the following manner before integration

$$A^B(x) = A^B \left[1 + Ce^{\left(\frac{-2\gamma_0|x+(i-1)L_e^B - l_c|}{w^B}\right)}\right]^{-1} \tag{16}$$

where i ∈ [1,2,3 ... n] where n is the total number of elements in the blade. The torsional stiffness of the blade with a crack is approximated as

$$GJ_{X'}^{B}(x) = GJ_{X'}^{B}\left[1 + Ce^{\left(\frac{-2\gamma_0|x-l_C|}{d}\right)}\right]^{-1}, C = \frac{J_{X'}^{B}-J_{X'}^{B_C}}{J_{X'}^{B_C}}, J_{X'}^{B} = I_{Y'}^{B} + I_{Z'}^{B}, J_{X'}^{B_C} = I_{Y'}^{B_C} + I_{Z'}^{B_C} \quad (17)$$

The functions $J_{X'}^{B}(x)$ is defined in the domain $\in [0, L^B]$ where $L^B$ is the total length of the blade, and the shape functions are defined in the domain $\in [0, L_e^B]$ where $L_e^B$ is the length of one element of the blade. Therefore, the coordinate system of the function $J_{X'}^{B}(x)$ has to be shifted for each element in the following manner before integration

$$J_{X'}^{B}(x) = J_{X'}^{B}\left[1 + Ce^{\left(\frac{-2\gamma_0|x+(i-1)L_e^B-l_C|}{w^B}\right)}\right]^{-1} \quad (18)$$

where i ∈ [1,2,3 ... n] where n is the total number of elements in the blade.

2.1.2. Construction of Force vector

The forces acting on the blades are aerodynamic and inertial. The aerodynamic forces are generated because of the aerofoil shape of the turbine blade. The aerofoil shape gives rise to lift and drag forces, and the rotation exerts a centrifugal force on the blades. These forces are then used to construct the force vector (Figure 1(c)) using the appropriate finite element approximation scheme. The total aerodynamic lift, drag, and centrifugal forces [33] acting of an individual blade can be approximated as

$$\bar{F}_C = \int_r^{L^B+d^D/2} \rho_b A^B \omega^2 r \, dr \quad (19)$$

$$\bar{F}_L = \int_{d/2}^{L^B+d^D/2} 0.5 \rho_a V_\alpha^2 C_L S \, dr \quad (20)$$

$$\bar{F}_D = \int_{d/2}^{L^B+d^D/2} 0.5 \rho_a V_\alpha^2 C_D S \, dr \quad (21)$$

$$V_\alpha^2 \cong (r\omega)^2 + V_\infty^2 \quad (22)$$

where $\bar{F}_C$ is the total centrifugal force, $\bar{F}_L$ is the total lift force, $\bar{F}_L$ is the total drag force, $\rho_b$ is blade mass density, $A^B$ is the area of the cross-section along the length of the blade, $\omega$ is the angular speed of the shaft, $d^D$ is the disc diameter, $L^B$ is the length of the blade, $\rho_a$ is the density of air, $V_\infty$ is the free stream velocity, S is the chord length at a given cross-section, $C_L$ is the coefficient of lift and $C_D$ is the coefficient of drag. If $\varepsilon_b$ is the angle of downwash of blades, the forces along the blade coordinate system can be written as

$$F_{X'} = \bar{F}_C \quad (23)$$

$$F_{Y'} = -\bar{F}_D \cos \varepsilon_b - \bar{F}_L \sin \varepsilon_b \quad (24)$$

$$F_{Z'} = \bar{F}_L \cos \varepsilon_b - \bar{F}_D \sin \varepsilon_b \tag{25}$$

The force along the X direction can be calculated by using the shape function and acceleration produced by centrifugal force. However, the shape function is defined only in the domain $\in [0, L_e^B]$ where $L_e^B$ is the length of an element of the blade, and the centrifugal force is defined in the domain of the complete blade length, which is $\in [0, L^B]$, we have shifted the coordinate system of the centrifugal force appropriately for each blade before integration, as shown in the following equation

$$(F_X)_j = \int_0^{L_e^B} \rho_a A_b N_j \left(x + (i-1)L_e^B + \frac{d^D}{2}\right) \omega^2 dx \tag{26}$$

for $j \in [1,2]$ where $N_1 = \left(1 - \frac{x}{L_e^B}\right), N_2 = \left(1 + \frac{x}{L_e^B}\right)$ and $i \in [1,2,3 \dots n]$ where n is the total number of elements in the blade. Using a similar approach along Y direction, we can define the forces as follows

$$(F_Y)_j = \int_0^{L_e^B} \rho_a A_b N_j (-\bar{F}_D \cos \varepsilon_b - \bar{F}_L \sin \varepsilon_b) dx \tag{27}$$

for $j \in [1,2]$ where $N_1 = \left(1 - 3\frac{x^2}{L_e^{B2}} + 2\frac{x^3}{L_e^{B3}}\right), N_2 = \left(\frac{3x^2}{L_e^{B2}} - \frac{2x^3}{L_e^{B3}}\right)$.

$$(F_Z)_j = \int_0^{L_e^B} \rho_a A_b N_j (\bar{F}_L \cos \varepsilon_b - \bar{F}_D \sin \varepsilon_b) dx \tag{28}$$

for $j \in [1,2]$ where $N_1 = \left(1 - 3\frac{x^2}{L_e^{B2}} + 2\frac{x^3}{L_e^{B3}}\right), N_2 = \left(\frac{3x^2}{L_e^{B2}} - \frac{2x^3}{L_e^{B3}}\right)$. These forces also create corresponding moments which are shown in Appendix-A

Now the whole force vector can be populated using the terms calculated above. The coordinate system of every blade is different from the global coordinate system. Hence, the system matrices for each blade have to be transformed before assembling them into the global system matrices. The absolute angular position of the j[th] blade with respect to the global coordinate system is determined as follows:

1) When t<0.2 (Figure 1(d)), that is when the angular velocity is increasing, the angular position can be calculated as

$$\theta_j^T = \theta^S + \theta_j^B, \quad \theta^S = \omega t/2, \theta_j^B = \frac{(j-1)2\pi}{N} \tag{29}$$

2) When t>0.2 (Figure 1(d)) that is when the angular velocity is constant, the angular position can be calculated as

$$\theta_j^T = \theta^S + \theta_j^B, \quad \theta^S = \frac{0.2}{2}\omega + \omega(t - 0.2), \theta_j^B = \frac{(j-1)2\pi}{N} \tag{30}$$

where $\theta^S$ is due to shaft rotation and $\theta_j^B$ is the angular position of the j[th] blade with respect to the shaft coordinate system, $\omega$ is the rotational speed of the shaft. The full transformation matrix for the j[th] blade can now be calculated as

$$T^f = \begin{bmatrix} T_j & 0 & 0 & 0 \\ 0 & T_j & 0 & 0 \\ 0 & 0 & T_j & 0 \\ 0 & 0 & 0 & T_j \end{bmatrix}, where\ T_j = \begin{bmatrix} \cos\left(\frac{\pi}{2}\right) & \cos\left(\frac{\pi}{2}+\theta_j{}^T\right) & \cos(\theta_j{}^T) \\ \cos\left(\frac{\pi}{2}\right) & \cos(\theta_j{}^T) & \cos\left(\frac{\pi}{2}-\theta_j{}^T\right) \\ \cos(\pi) & \cos\left(\frac{\pi}{2}\right) & \cos\left(\frac{\pi}{2}\right) \end{bmatrix} \quad (31)$$

The blade system equation is transformed into the global coordinate system using standard coordinate transformation. The details are mentioned in Appendix-A.

2.2. Simulation Scheme

The finite element system Eq. (1) is discretized with time steps with the help of Newmark's time integration method [34],

$$\dot{u}_{t+\Delta t} = \dot{u}_t + (1-\alpha)\Delta t \ddot{u}_t + \alpha \Delta t \ddot{u}_{t+\Delta t} \quad (32)$$

$$u_{t+\Delta t} = u_t + \Delta t \dot{u}_t + (0.5-\beta)\Delta t^2 \ddot{u}_t + \beta \Delta t^2 \ddot{u}_{t+\Delta t} \quad (33)$$

where $\alpha = 1/2$ and $\beta = 1/4$ give unconditional stability for a linear system and are applicable for a linear time-varying system with limit-cycle oscillation as in the present case due to the periodic rotational position of the blades with sufficiently fine time steps within the rotational time period. Substituting $\ddot{u}_{t+\Delta t}$ and $\dot{u}_{t+\Delta t}$ in terms of $u_{t+\Delta t}$ from Eq. (32) and (33) in Eq. (1) and rearranging, one has

$$\left(K + \frac{1}{\beta \Delta t^2}M + \frac{\alpha}{\beta \Delta t}C\right)u_{t+\Delta t} = f_{t+\Delta t} + M\{\frac{1}{\beta \Delta t^2}u_t + \frac{1}{\beta \Delta t}\dot{u}_t + (\frac{1}{2\beta}-1)\ddot{u}_t\} + C\{\frac{\alpha}{\beta \Delta t}u_t + \left(\frac{\alpha}{\beta}-1\right)\dot{u}_t + \frac{\Delta t}{2}\left(\frac{\alpha}{\beta}-2\right)\ddot{u}_t\} \quad (34)$$

which gives the final reduced form to be solved at each time step as

$$\bar{K}u_{t+\Delta t} = \bar{f} \quad (35)$$

where $\bar{K}$ is the effective or reduced stiffness matrix for the time step and the reduced force vector $\bar{f}$ is obtained for the time step in terms of the variables from the previous time steps.

2.3. Numerical Convergence Study

In the next step of our study, we aim to determine the optimum number of finite elements in discretized model. To compare the effect of increasing number of elements, we plot the Fourier transform of shaft tip vibration in Z direction with respect to time.

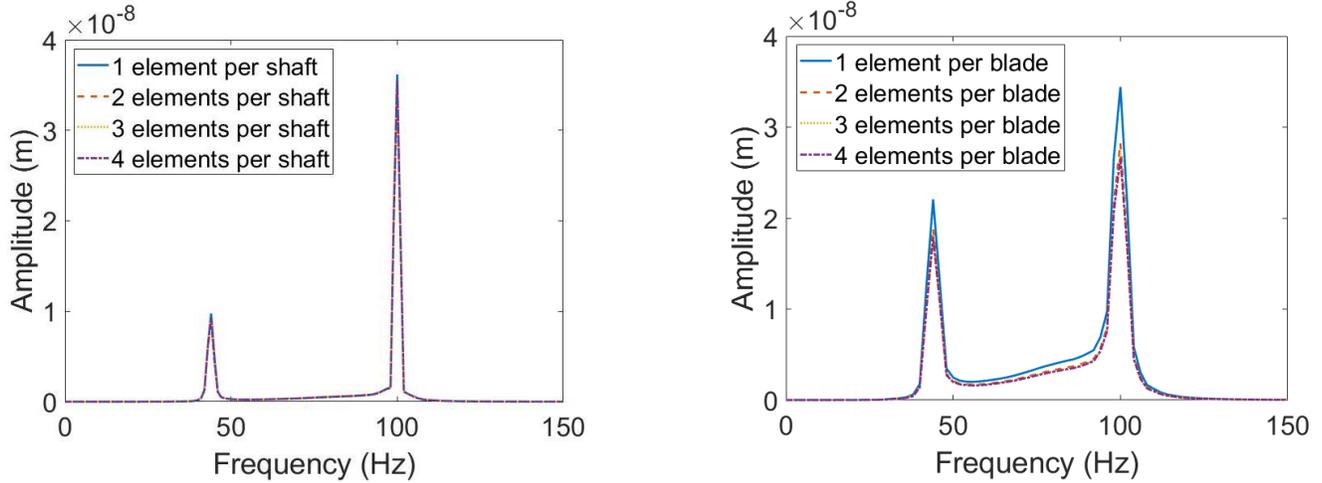

Figure 3(a) Effect of increasing the number of finite elements in the shaft on shaft tip radial displacement. (b) Effect of increasing the number of elements in the blade shaft tip radial displacement.

Figure 3(a) shows that an increasing number of elements in the shaft do not affect the FFT amplitude and the fundamental frequencies. The first peak corresponds to a local resonance and the second peak corresponds to the forcing frequency. Similarly, increasing the number of elements per blade creates a minor change in the amplitude (Figure 3(b)) of the FFT plot while there is no change in the fundamental frequencies. Based on this observation, we have used 1 element in the shaft and 2 elements per blade for further study. In the first part of the study, the model results have been compared with some experimental data for validation, followed by three case studies, each dealing with a particular case of damage.

2.4. Model validation

The reduced-order model of the bladed disk system is first validated against experimental results [35]. For comparing the dynamic response of the blade model with the experimental data, the first natural frequency was compared for healthy and damaged states. The experimental setup consists of exciting a blade with a frequency sweep carried out with a shaker and measuring vibration signals at the blade tip (Figure 4). The vibration signals were measured using a Laser Doppler Velocimetry (LDV). The vibration signals were then transformed into the frequency domain by Fourier transform, and the resonances were determined from the dominant peaks, and the corresponding natural frequencies were determined. For the proposed model, the natural frequency ω was calculated by solving the eigenvalue problem.

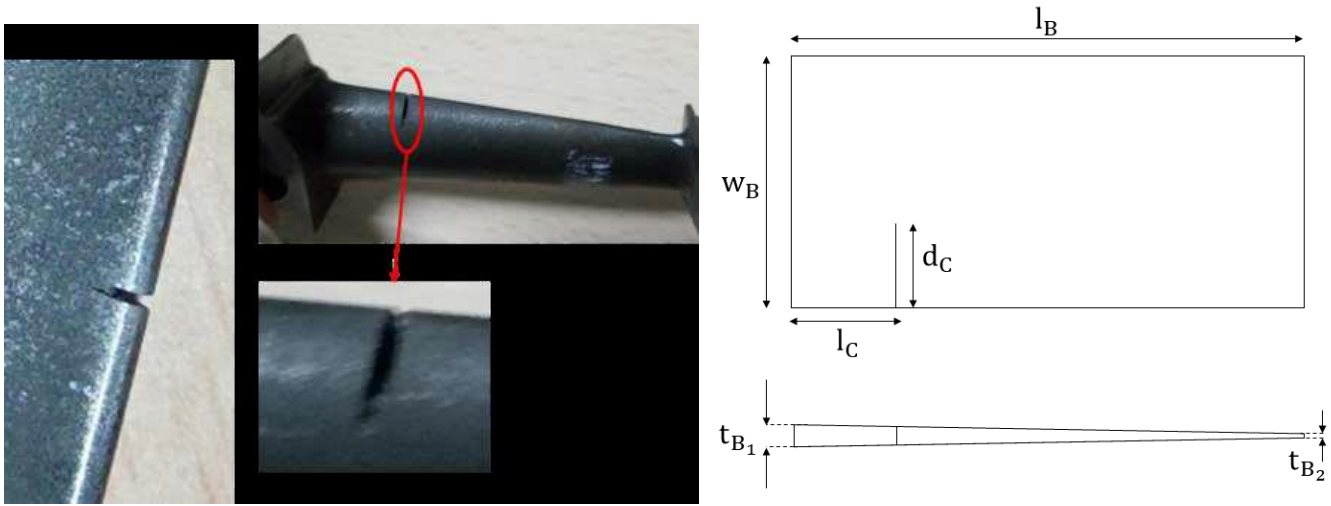

Figure 4 (a) Blade with crack with an edge crack on the leading edge (b) Top and side view of the blade showing the relevant parameters. The crack length used for comparison of experimental and simulated data were 1.5mm and 2.5mm on the leading edge. [33].

Table 1 Comparison of simulated results with experimental results

| Damage Case | First natural frequency (experimental) | Change in natural frequency | First natural frequency (Model) | Change in natural frequency |
|---|---|---|---|---|
| **Undamaged** | 2508 | - | 2504.7 | - |
| **1.5mm crack** | 2471 | 1.47% | 2470.4 | 1.36% |
| **2.5mm crack** | 2431.6 | 3.04% | 2446.8 | 2.3% |

For simulating the damage in the blade, a single crack of specific length and location as the experiment was modelled as described in the previous section. Three cases were considered for comparison, as shown in Table 1. The first case was the undamaged blade, the second case was a blade with 1.5mm edge crack and the final case was 2.5mm edge crack (Figure 4(a)). Furthermore, the change from the undamaged blade's natural frequency was also calculated to see the impact of the presence of the crack in the blade. It was found that the simulated results are in good agreement with the experimental results. The results calculated from the simulation and the experiment are shown in Table 1. After the model was validated with the experimental data, three damage case studies were carried out, which are discussed in the next section.

Table 2 System parameters

| System Property | Value | System Property | Value |
|---|---|---|---|
| **No. of Blades** | 8 | Modulus (**E**) | 2e10$^9$ Pa |
| **Shaft outer diameter (d$^{Sout}$)** | 0.025 m | Fan Disk Density ($\rho^D$) | 4430 Kg/m$^3$ |
| **Shaft inner diameter (d$^{Sin}$)** | 0.015 m | Fan Blade Density ($\rho$) | 7833 Kg/m$^3$ |
| **Turbine shaft length of the first stage (L$^S$)** | 0.5 m | Shaft Density ($\rho$) | 7833 Kg/m$^3$ |
| **Turbine shaft length of the second stage (L$^S$)** | 0.5 m | Poisson's Ratio ($\vartheta$) | 0.31 |
| **Fan blade disk outer diameter (d$^D$)** | 0.35 m | Modulus (**G**) | 4.2e10$^{10}$ Pa |
| **Blade width (w$^B$)** | 0.04 m | Density of air ($\rho_a$) | 1.22 Kg/m$^3$ |
| **Blade thickness (t$^{B_1}$)** | 0.00515 m | Coefficient of lift ($C_L$) | 0.02 |
| **Blade thickness (t$^{B_2}$)** | 0.00065 m | Coefficient of drag ($C_D$) | 0.03 |
| **Blade Length (L$^B$)** | 0.40 m | Angle of attack of blade ($\varepsilon$) | 0.3 |
| **Crack depth (d$_C$)** | 0.0025 m | Free Stream air velocity ($V_\infty$) | 200 m/s |
| **Crack location (l$_C$)** | 0.025 m | Engine RPM ($\omega$) | 6000 |

### 3. Case Studies

To demonstrate the capability of the proposed reduced-order model, we report here three case studies. These are (1) cracks in the blade where an impact is applied on different blades, and the severity of the crack on the system response is studied, including Fan Blade Off (FBO), (2) Foreign Object Damage (FOD), and (3) cracks in a two-stage bladed disk. The system parameters taken for the following case studies are listed in Table 2.

3.1. Case study 1: Crack in the blade

Mistuning in the bladed disk is caused when the properties of one of the blades are different from the others. We assume a single crack in one or multiple blade(s), as shown in Figure 5(a). The angular velocity of the shaft was increased from 0 to 6000 rpm in 0.2 sec with a linear ramp followed by a constant speed, as shown in Figure 5(b). The shaft tip radial displacement time history is analyzed to capture the effect of the damage on the bladed disk system. The radial displacement was calculated as

$$u_r = \sqrt{u_y^2 + u_z^2} \tag{36}$$

The shaft tip radial displacement in the presence of a crack of 10mm length located at 10mm from the blade root is shown in Figure 5(c). Furthermore, it can be seen from Figure 5(d) that there is a local resonance happening with one of the natural frequencies of the full assembly (Table 3) as RPM was increasing through the blade resonance frequency of around 70Hz. The phenomenon is captured well by the reduced-order model. For the study of the effect of crack length on the shaft tip radial displacement, different crack lengths of 5mm, 7.5mm, 10mm and 12.5mm located at 100mm from the blade root were considered. The radial displacement of the shaft tip with respect to time is shown in Figure 6(a). It can be seen from the graph that the crack size of length 5mm causes the lowest radial displacement of the shaft tip and the crack size of 12.5mm causes the highest radial displacement of the shaft tip. The increase in the radial displacement of the shaft indicates that there is an increase in the imbalance in the system because of the increase in the crack length. The effect of the location of a crack along the length of the blade was also simulated for four crack locations at 10mm, 25mm, 50mm, and 75mm from the root of the blade with a crack length of 10mm. It can be seen from Figure 6(b) that the crack located at 10mm from the blade root causes the highest radial displacement of the shaft tip and the crack at 75mm from the blade root causes the lowest radial displacement of the shaft tip.

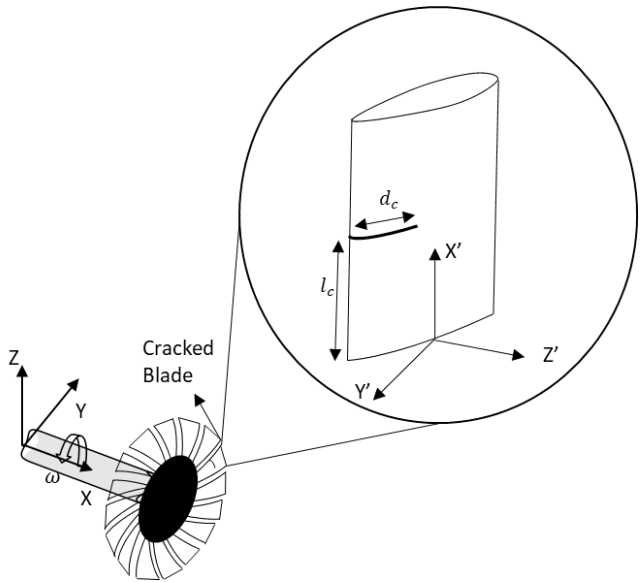
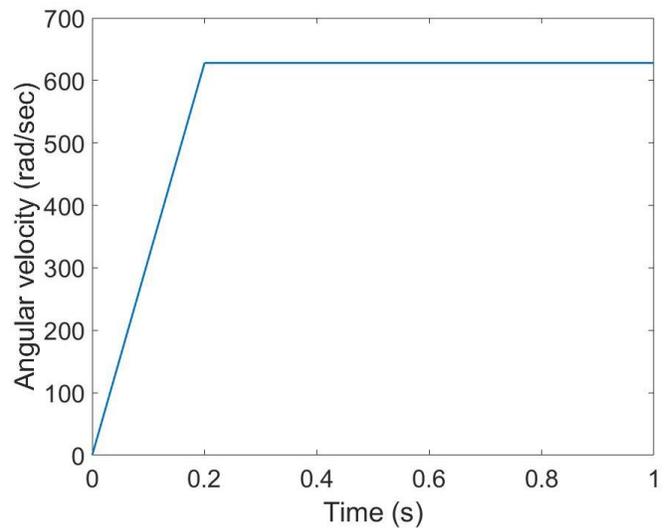

(a)          (b)

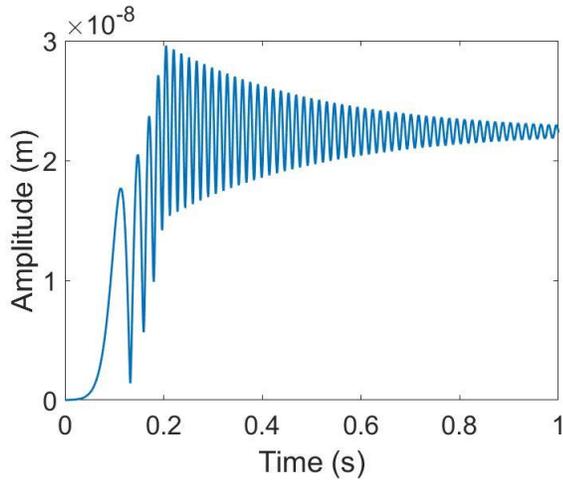
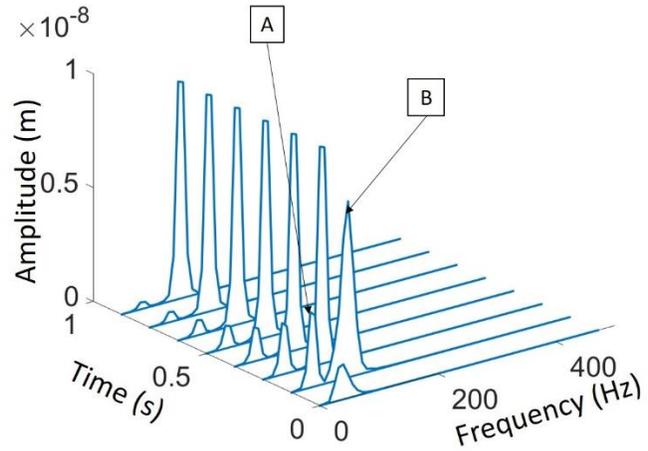

(c)            (d)

Figure 5 (a) Cracked blade in the bladed disk system (b) The angular velocity of the shaft with respect to time (c) Radial displacement of shaft tip for a crack of 10mm located at 10mm from the blade root. (d) Moving time-window FFT plot of shaft tip displacement for blade with a crack. Label A shows the peaks corresponding to local resonance when the shaft is speeding up, and Label B shows the peak corresponding to the excited frequency because of the system rotation.

Table 3 First five natural frequencies of the individual components and the whole assembly for both the damaged and undamaged cases.

| Natural Frequencies | Damaged Blade (Crack 10mm) | Undamaged Blade | Shaft | Full Assembly (Undamaged) | Full Assembly (Damaged, crack 10mm) |
|---|---|---|---|---|---|
| 1 | 2246.9 | 2505.6 | 389.4 | 70.37 | 70.37 |
| 2 | 14161.1 | 15828.1 | 389.4 | 676.93 | 676.81 |
| 3 | 17257.6 | 24412.9 | 3837.2 | 676.93 | 676.92 |
| 4 | 35387.7 | 50046.8 | 3837.2 | 1038.54 | 1035.93 |
| 5 | 47834.6 | 53533.5 | 8069.1 | 2505.62 | 2314.44 |

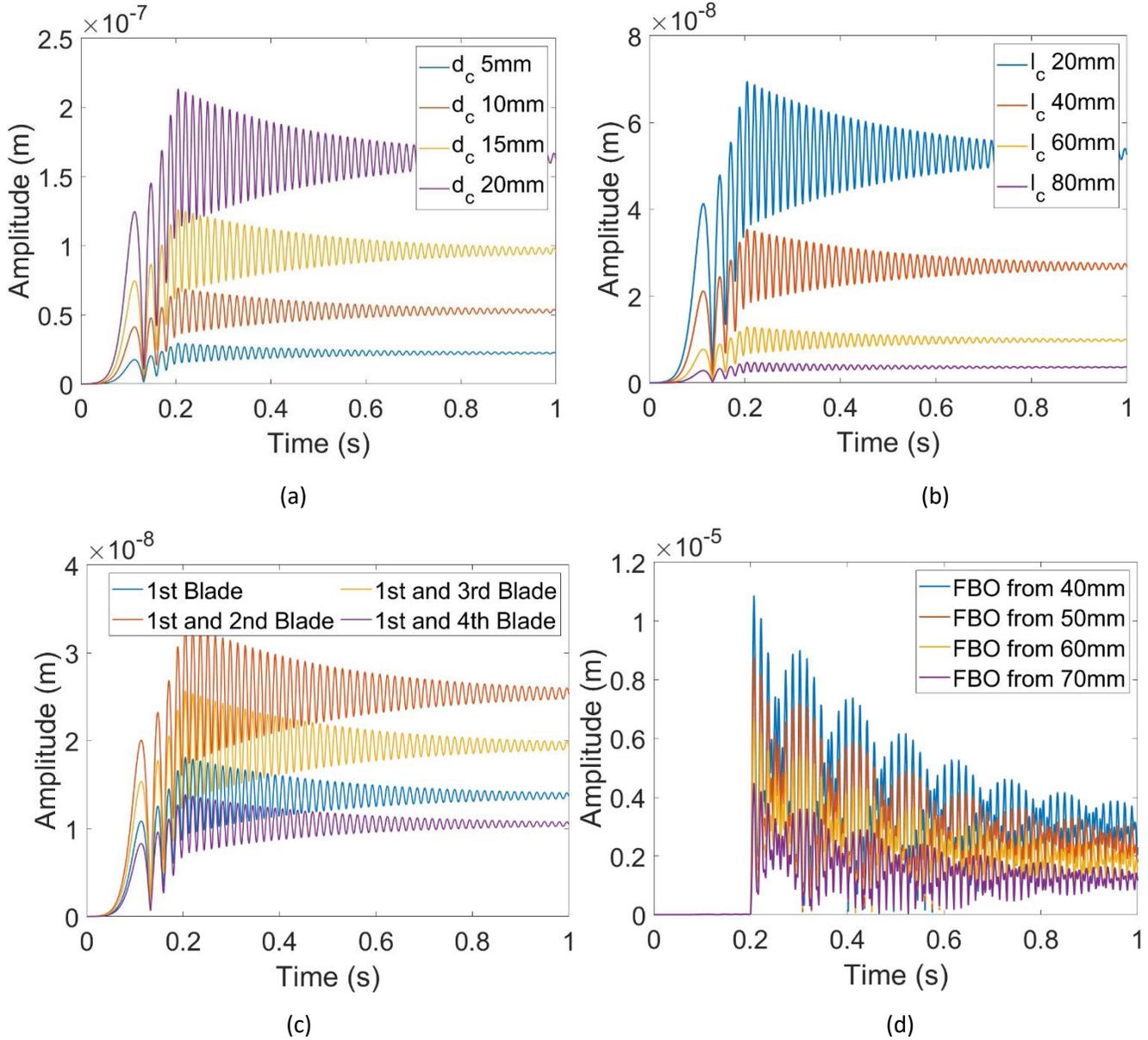

Figure 6 shows the (a) effect of crack length on shaft tip response shows that the increase in the crack length increases the radial displacement of the shaft tip. (b) Impact of crack location on shaft tip response. The crack located near the blade root gives rise to more radial displacement (more system imbalance) as compared to a crack located farther away from the root. (c) Impact of the relative position of the cracked blade on the system response. Two cracks located nearest to each other have the most significant effect on the system imbalance. (d) Impact of Fan blade off (FBO) on the shaft tip response for varying locations from the blade root. For FBO closest to the blade root, the imbalance is the highest in the system, thus giving rise to more radial shaft tip displacement.

The effect of the relative positions of two cracked blades is also studied. For this purpose, a single crack of length 5mm located at 75mm in the 1st blade is compared with a crack of the same characteristics (length and location) in (a) 1st and the 2nd blade, (b) 1st blade, and the 3rd blade and (c) 1st blade and the 4th blade. The system responsible for these four cases was compared with each other (Figure 6(c)). The presence of a crack in the adjacent blades has the most significant effect on the radial deflection of the shaft. The deflection reduced as the position of the cracked blade shifted farther away from each other. The cracked blades in the mutually opposite locations negate the imbalance created by each other. Also interesting to note is that the radial displacement of the shaft with the cracks present on diametrically opposite ends (1st and 4th blade) was less compared to the radial displacement due to a single crack on the 5th blade.

Fan blade off (FBO) event is likely when the crack becomes large, spanning over the width of the blade and a piece of blade comes off. Yu et. al.[30] calculated the shaft response for an FBO event for both the cases with and without the blade rubbing into the casing. Bladed disk response was studied for various values of inertial asymmetry. In our study, to model the asymmetry, the FBO event was simulated for a blade breaking from different lengths. The effect of the blade rubbing into the casing was not considered in the present study. However, such a situation can be simulated using constrained tip loads on the blade within the current framework of reduced-order modeling. Our model was able to capture the system response correctly (Figure 6(d)). Shaft displacement is more pronounced when the FBO event occurs closer to the root.

3.2. Case Study 2: Foreign object damage

A case study for a Foreign object damage (FOD) event was carried out on the bladed disk system. For simulating the collision of a foreign object, the impact force for a mass of 0.1Kg traveling at 25m/s was calculated, and the contact time was 0.04 seconds. The impact force was calculated as

$$F_{impact} = \frac{v_{impact}*mass}{t_{contact}} \tag{37}$$

The impact force was applied to the blade tip (Figure 7(a)) along the shaft axis. The shaft tip radial displacement response as a function of time was plotted as shown in Figure 7(b). For validating the FOD response, the present model results were compared with the result reported by Truso et al. [28]. An equivalent moment was applied at the shaft end for comparison of the effect of impact force since individual blades were not modelled in ref. [28]. However, in the present reduced-order model, the impact force is directly applied at the blade end, which is more realistic. The system imbalance was also introduced into the present system model by changing the length of one of the blades by 1mm for the purpose of comparison with the results from ref. [28]. The present model results are in good agreement. Figure 7(c) shows the shaft radial displacement for a FOD event with a varying magnitude of impact force. The FOD event causes a large transient response in the system, which eventually reduces and stabilizes because of the damping in the system. Furthermore, a case of FOD followed by the corresponding impacting mass sticking at the point of impact was also studied. Table 4 shows the magnitude of the impact force applied along with the mass stuck. It was observed that when the FOD event is followed by mass sticking, the imbalance created in the system is much larger. The radial displacement reduces to a steady-state response dependent on the extent of the imbalance created due to the mass stuck (Figure 7(d)). Figure 8(a) shows the moving time-window FFT plot for a FOD without mass sticking. The FOD event occurs at 0.2 seconds, because of which it can be observed that there are peaks around 30Hz-50Hz whose amplitude slowly decreases with time due to damping present in the system. The only visible peaks in the steady-state are at 100Hz because of a slight system imbalance, which is inherently present in the system.

Table 4  Impact force and mass stuck for each case

| Impact Case | Mass | Impact Force |
| --- | --- | --- |
| **Case 1** | 0.05Kg | 30N |
| **Case 2** | 0.1Kg | 60N |
| **Case 3** | 0.2Kg | 120N |
| **Case 4** | 0.5Kg | 310N |

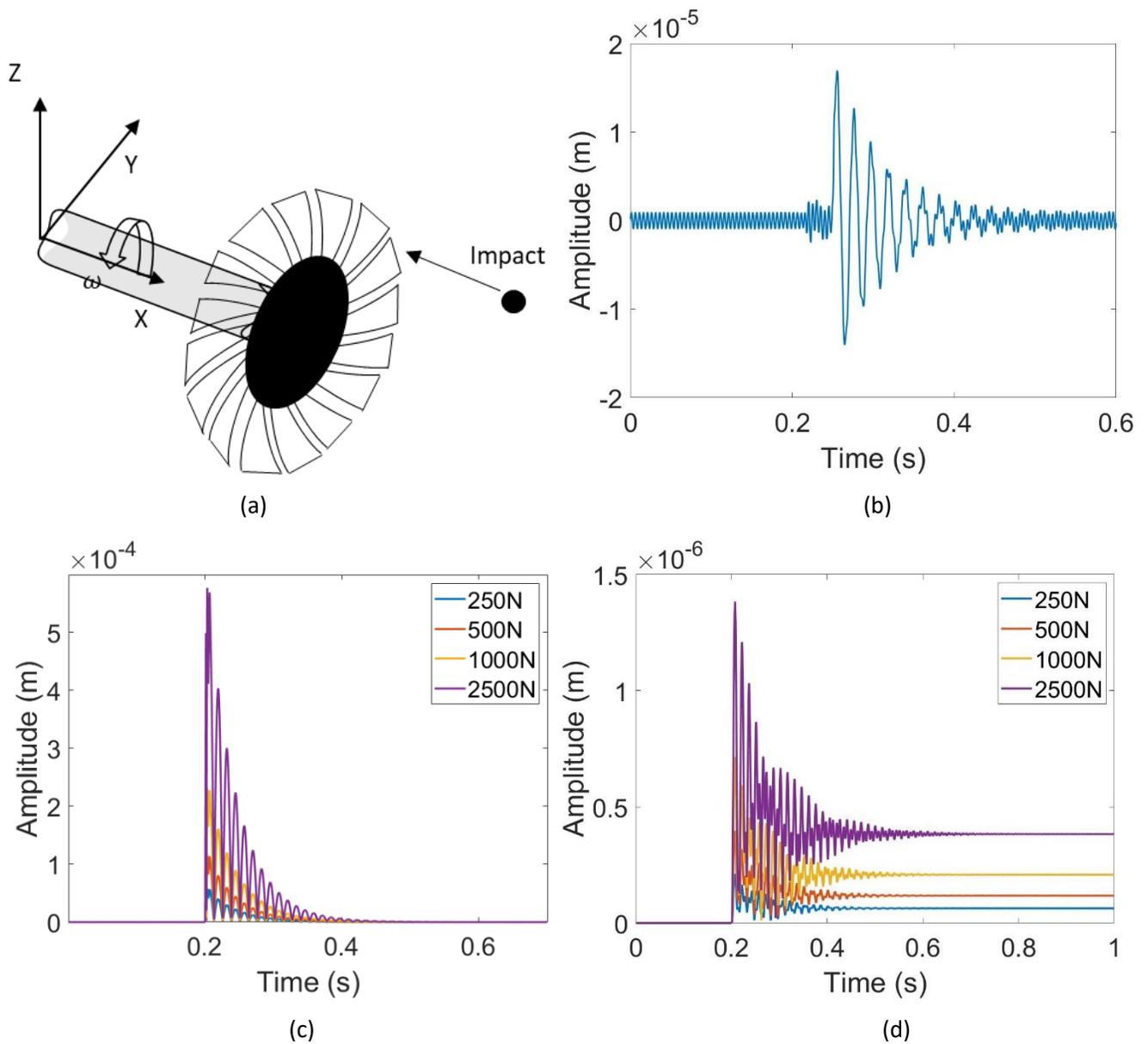

Figure 7 (a) Schematic diagram of foreign object damage event (b) Shaft tip displacement response along Z direction for 250N impact (c) Shaft tip radial response with increasing impact forces shows increasing impact forces creates higher radial shaft tip displacement (d) Shaft tip radial displacement response with increasing mass of impactor stuck to the blade showing that the increase of mass increases the shaft tip radial displacement.

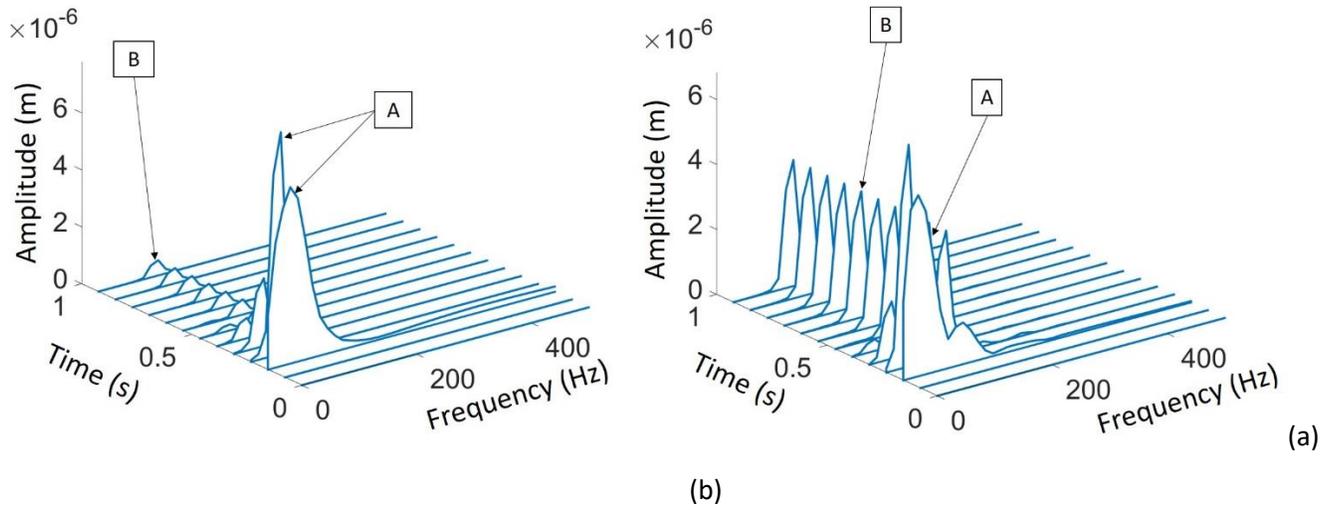

Figure 8 Moving time-window short-time FFT plot for (a) FOD event without mass stick. Label A shows the peaks corresponding to the instantaneous excited frequencies at 0.2 seconds when the impact happens. Label B shows peaks corresponding to the steady state excitation due to system rotation. (b) FOD event with mass sticking to the blade. Label A shows the peak corresponding to 0.2 seconds when the impactor mass sticks to the fan blade. Label B shows the peaks corresponding to the steady state excitation because of system rotation. The amplitude of the peaks is higher as compared to FOD without mass sticking because of higher level of imbalance created due to mass sticking.

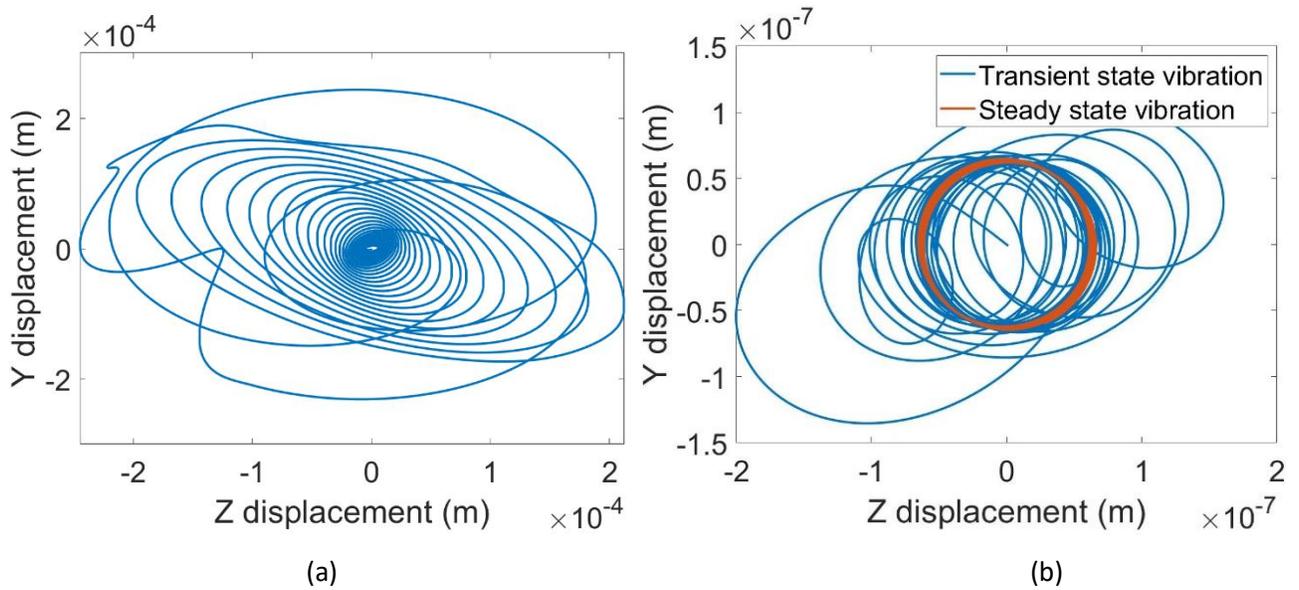

Figure 9 Shaft tip displacement trajectory for (a) FOD event without impactor mass sticking to the blade (b) FOD event with impactor mass sticking to the blade.

Figure 8(b) shows a similar moving time-window short-time FFT plot for FOD followed by mass sticking. In this case, the dominant peak occurs at 100Hz (which is the system RPM) at around 0.2 seconds, indicating the presence of imbalance present in the system due to extra mass stuck to the blade. Figure 9 shows the path followed by the shaft tip for FOD with and without mass sticking. It is evident from Figure 9(a) that the impact force due to the FOD event causes a large deflection in the shaft tip. This deflection slowly attenuates to the initial position (zero radial deflection) because of the damping present in the system. However, the imbalance caused by the FOD event with a mass sticking to the blade creates a permanent imbalance in the system. In the steady-state, the system revolves with a constant radius about its axis because of the permanent imbalance caused by the mass sticking (Figure 9(b)).

3.3. Case Study 3: Two-stage bladed disk

Moving to a more complex case, we consider a two-stage bladed disk and simulate cracks and FBO using the reduced-order modelling framework. A single crack of length 10mm at 10mm from the root was first introduced in the first stage and then in the second stage. The shaft radial displacement was studied for both disks.

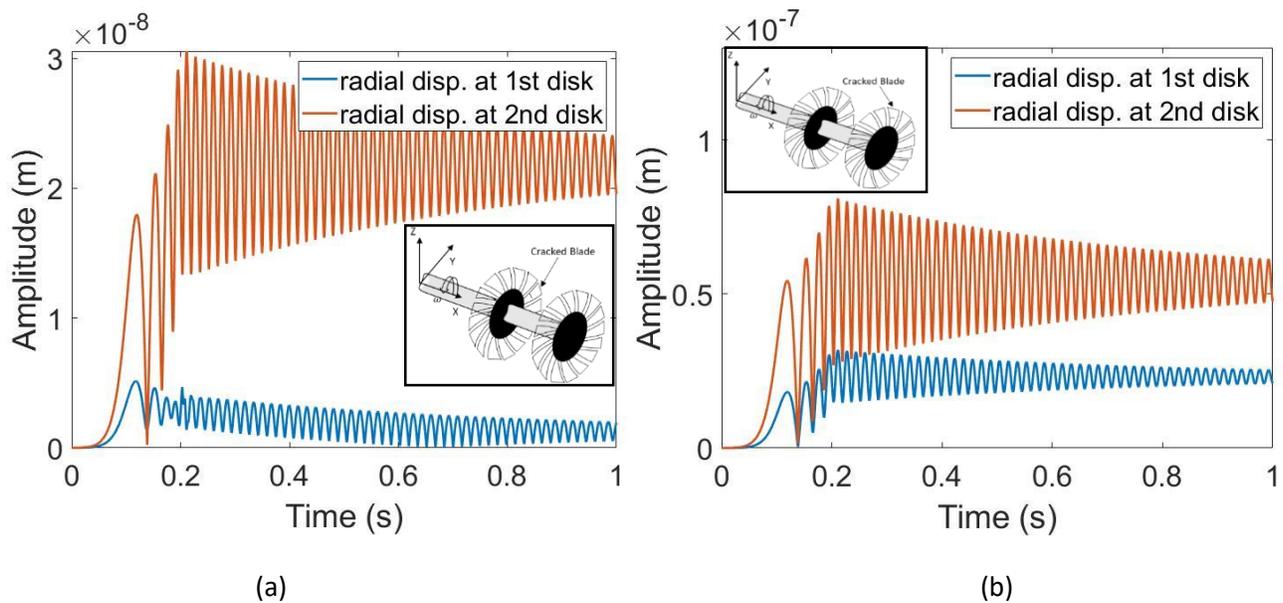

(a)  (b)

Figure 10 Shaft radial displacement of both stages for (a) Crack in 1st stage (b) Crack in 2nd stage showing that the displacement is more in the second stage when the crack is present either in the first stage or in the second stage

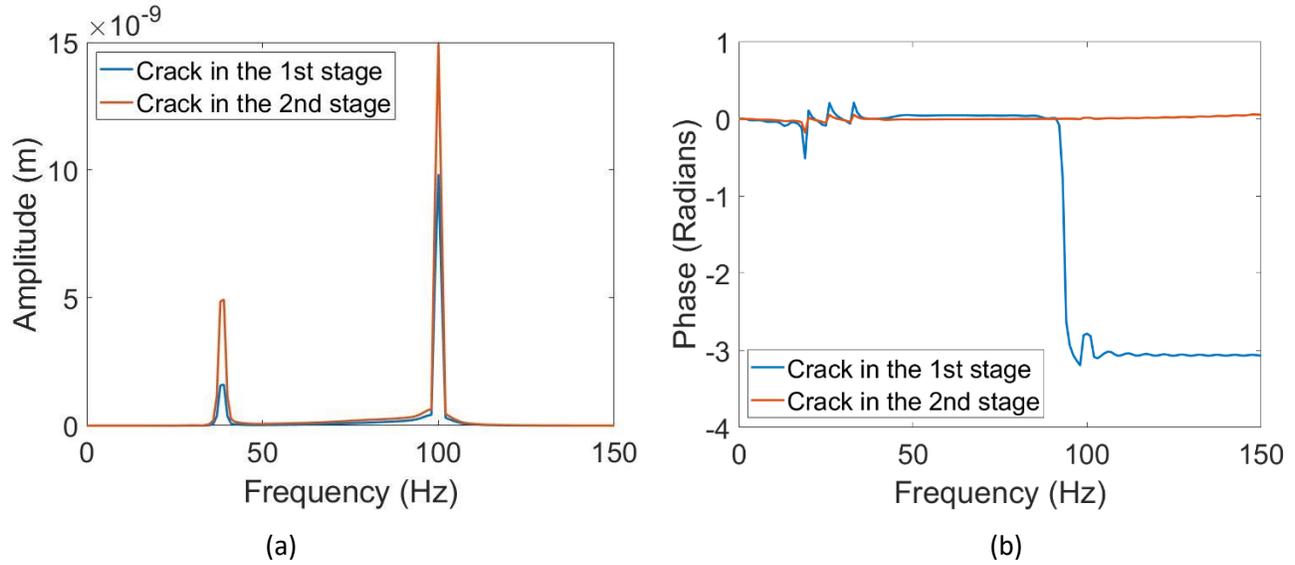

(a) (b)

Figure 11 (a) FFT of difference between displacement at 1st and 2nd disk location when the crack is present in the 1st stage vs when the crack is present in the 2nd stage (b) Phase of difference between displacement at 1st and 2nd disk location when the crack is present in the 1st stage vs when the crack is present in the 2nd stage.

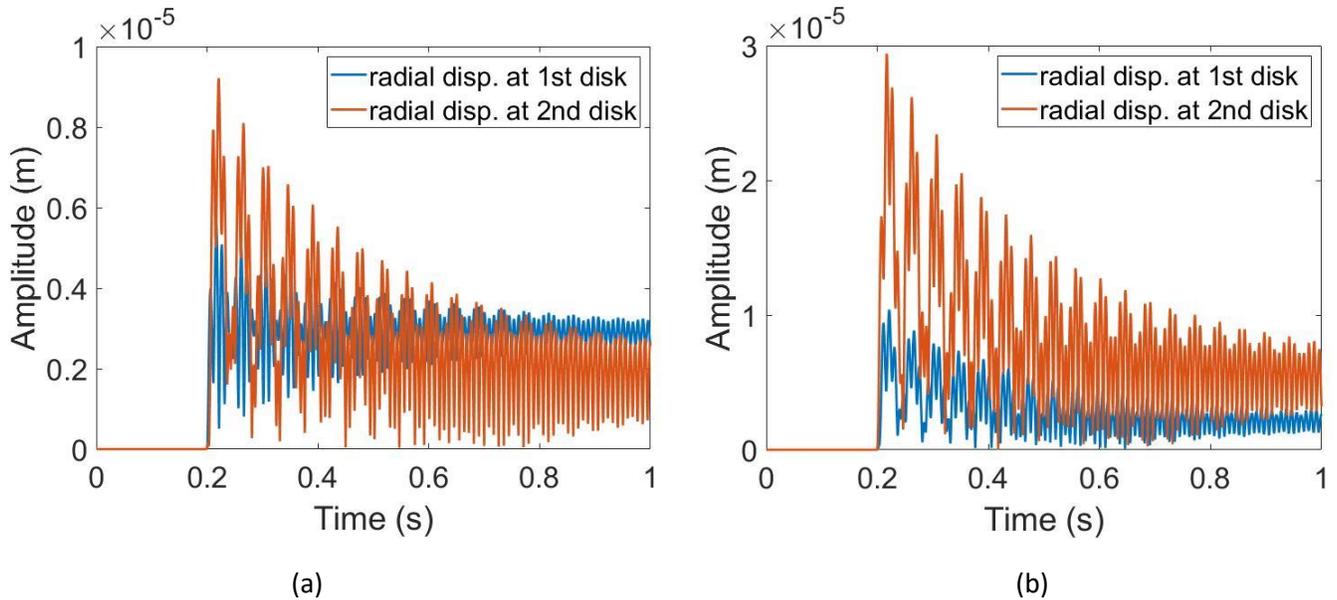

(a) (b)

Figure 12 Shaft tip radial displacement for both stages for (a) FBO in 1st Stage (b) FBO in 2nd Stage.

Figure 10 indicates that an imbalance is introduced in both the stages of the blades when a crack is present in either stage. However, the magnitude of the imbalance is more in the second stage as compared to the first stage for both cases. Figure 11 has been plotted by calculating FFT amplitude of the difference between the radial displacement of stage 1 and 2 at the disk-shaft locations when the crack is present in stage 1 and when the crack is present in the stage 2. Mathematically it can be represented in the following way

$$U_1 = \mathcal{F}((u_r^{\text{Stage 2}}) - (u_r^{\text{Stage 1}})_{\text{Crack Stage 1}}) \qquad (38)$$

$$U_2 = \mathcal{F}((u_r^{\text{Stage 2}}) - (u_r^{\text{Stage 1}})_{\text{Crack Stage 2}}) \qquad (39)$$

where $\mathcal{F}$ represents the Fourier transform of the time signal. Figure 11(a) was generated by calculating the amplitude of the complex quantities $U_1$ and $U_2$. Similarly, Figure 11(b) was generated by calculating the phase angle of the complex quantities $U_1$ and $U_2$. The peaks in Figure 11(a) indicate that the vibration amplitude of the second stage is more than the first stage. Figure 11(b) shows that the presence of crack also induces a phase difference in the vibration between the two stages.

The FBO event was simulated in a similar fashion with FBO in the first stage and then in the second stage. For both cases, the blade fracture was introduced at 50mm from the root. The vibration amplitude for both stages was increases abruptly when the FBO event occurred in the first stage. The maximum amplitude of the second stage was found to be more than the first stage at the event of blade off (Figure 12(a)) initially. However, the vibration of the second stage slowly reduces to a level below the first stage vibration when the steady-state is reached. When FBO event occurred in the second stage, there was an abrupt increase in the vibration amplitude of both stages (Figure 12(b)). However, initially, the vibration amplitude of the second stage was almost three times greater than the first stage. In the steady-state, the FBO event in the second stage gave rise to a higher level of vibration in the second stage shaft tip than in the first stage.

4. Conclusion

A reduced-order modeling framework for a bladed disk system is proposed in the paper. In order to model the system in 3 dimensions, two bending directions and an extension direction were considered for each blade besides rotational degrees of freedom. The stiffness matrix and consistent mass matrix of the bending and extension were calculated from finite element beam formulations. Rayleigh damping was used to construct the damping matrix. The aerodynamic, centrifugal forces, and moments acting on the blade were calculated as the consistent force vector on the finite element nodes. Three cases of damages viz. blade mistuning due to crack presence, foreign object damage, and fan blade off were modeled. Investigations were performed for both single staged and two-staged bladed disks. The proposed reduced-order model was able to capture the effects of the complex interaction of damages. First, a convergence study was carried out to find the optimum number of elements per blade and shaft. Then, the ability of the proposed model to simulate damage signature in the form of vibration response changes was validated with experimental results. Fan blade off and foreign object damage simulation results were verified with studies from the literature. The current reduced-order framework of the bladed disk successfully simulates results up to a satisfactory degree of accuracy.

The effects of the crack length, location, and relative position of the crack on the blades were simulated. It was found that the imbalance increased with the increase in crack size and closeness to the blade root. In case of foreign object damage, the impact force creates a moment at the shaft end, which creates a radial deflection. For the fan blade off event, it was found that sudden change in the length of the blade creates a large abrupt transient in the shaft radial displacement. Finally, a two-stage bladed disk with coupled effects was modelled, and the effect of the presence of crack and FBO event in both the stages in different combinations were studied.  The reduced-order modelling framework proposed here will be useful to simulate probabilistically various parametric effects and identify their correlations to the possible sensing signals that can be obtained in real-time toward developing an integrated health monitoring system.

In future investigations, a more refined model of damages may be included. A more detailed model of crack and related nonlinearities in the finite element model can be taken to account for crack opening and closing. The simulated signals from such models could be used for training a deep learning model for classification, localization, and identification for online health monitoring applications.

**Appendix-A**

The individual elements of the turbine blade system is formulated as per the basic finite element method. The complete mass matrix of the shaft can be written as

$M_S =$

$$\begin{bmatrix}
\frac{\rho A_S L_S}{3} & 0 & 0 & 0 & 0 & 0 & \frac{\rho A_S L_S}{6} & 0 & 0 & 0 & 0 & 0 \\
0 & \frac{156\rho A_S L_S}{420} & 0 & 0 & 0 & \frac{22\rho A_S L_S^2}{420} & 0 & \frac{54\rho A_S L_S}{420} & 0 & 0 & 0 & \frac{-13\rho A_S L_S^2}{420} \\
0 & 0 & \frac{156\rho A_S L_S}{420} & 0 & \frac{22\rho A_S L_S^2}{420} & 0 & 0 & 0 & \frac{54\rho A_S L_S}{420} & 0 & \frac{-13\rho A_S L_S^2}{420} & 0 \\
0 & 0 & 0 & \frac{\rho J_S L_S}{3} & 0 & 0 & 0 & 0 & 0 & \frac{\rho J_S L_S}{6} & 0 & 0 \\
0 & 0 & \frac{22\rho A_S L_S^2}{420} & 0 & \frac{4\rho A_S L_S^3}{420} & 0 & 0 & 0 & \frac{13\rho A_S L_S^2}{420} & 0 & \frac{-3\rho A_S L_S^3}{420} & 0 \\
0 & \frac{22\rho A_S L_S^2}{420} & 0 & 0 & 0 & \frac{4\rho A_S L_S^3}{420} & 0 & \frac{13\rho A_S L_S^2}{420} & 0 & 0 & 0 & \frac{-3\rho A_S L_S^3}{420} \\
\frac{\rho A_S L_S}{6} & 0 & 0 & 0 & 0 & 0 & \frac{\rho A_S L_S}{3} & 0 & 0 & 0 & 0 & 0 \\
0 & \frac{54\rho A_S L_S}{420} & 0 & 0 & 0 & \frac{13\rho A_S L_S^2}{420} & 0 & \frac{156\rho A_S L_S}{420} & 0 & 0 & 0 & \frac{-22\rho A_S L_S^2}{420} \\
0 & 0 & \frac{54\rho A_S L_S}{420} & 0 & \frac{13\rho A_S L_S^2}{420} & 0 & 0 & 0 & \frac{156\rho A_S L_S}{420} & 0 & \frac{-22\rho A_S L_S^2}{420} & 0 \\
0 & 0 & 0 & \frac{\rho J_S L_S}{6} & 0 & 0 & 0 & 0 & 0 & \frac{\rho J_S L_S}{3} & 0 & 0 \\
0 & 0 & \frac{-13\rho A_S L_S^2}{420} & 0 & \frac{-3\rho A_S L_S^3}{420} & 0 & 0 & 0 & \frac{-22\rho A_S L_S^2}{420} & 0 & \frac{4\rho A_S L_S^3}{420} & 0 \\
0 & \frac{-13\rho A_S L_S^2}{420} & 0 & 0 & 0 & \frac{-3\rho A_S L_S^3}{420} & 0 & \frac{-22\rho A_S L_S^2}{420} & 0 & 0 & 0 & \frac{4\rho A_S L_S^3}{420}
\end{bmatrix}$$

(40)

Similarly, the complete stiffness and displacement matrix can be assembled as

$$K_S = \begin{bmatrix}
\frac{EA_S}{L_S} & 0 & 0 & 0 & 0 & 0 & \frac{-EA_S}{L_S} & 0 & 0 & 0 & 0 & 0 \\
0 & \frac{12EI_Z}{L_S^3} & 0 & 0 & 0 & \frac{6EI_Z}{L_S^2} & 0 & \frac{-12EI_Z}{L_S^3} & 0 & 0 & 0 & \frac{6EI_Z}{L_S^2} \\
0 & 0 & \frac{12EI_Y}{L_S^3} & 0 & \frac{-6EI_Y}{L_S^2} & 0 & 0 & 0 & \frac{-12EI_Y}{L_S^3} & 0 & \frac{-6EI_Y}{L_S^2} & 0 \\
0 & 0 & 0 & \frac{GJ_S}{L_S} & 0 & 0 & 0 & 0 & 0 & \frac{\rho J_S L_S}{6} & 0 & 0 \\
0 & 0 & \frac{-6EI_Y}{L_S^2} & 0 & \frac{4EI_Y}{L_S} & 0 & 0 & 0 & \frac{6EI_Y}{L_S^2} & 0 & \frac{2EI_Y}{L_S} & 0 \\
0 & \frac{6EI_Z}{L_S^2} & 0 & 0 & 0 & \frac{4EI_Z}{L_S} & 0 & \frac{-6EI_Z}{L_S^2} & 0 & 0 & 0 & \frac{2EI_Z}{L_S} \\
\frac{-EA_S}{L_S} & 0 & 0 & 0 & 0 & 0 & \frac{EA_S}{L_S} & 0 & 0 & 0 & 0 & 0 \\
0 & \frac{-12EI_Z}{L_S^3} & 0 & 0 & 0 & \frac{-6EI_Z}{L_S^2} & 0 & \frac{12EI_Z}{L_S^3} & 0 & 0 & 0 & \frac{-6EI_Z}{L_S^2} \\
0 & 0 & \frac{-12EI_Y}{L_S^3} & 0 & \frac{6EI_Y}{L_S^2} & 0 & 0 & 0 & \frac{12EI_Y}{L_S^3} & 0 & \frac{6EI_Y}{L_S^2} & 0 \\
0 & 0 & 0 & \frac{\rho J_S L_S}{6} & 0 & 0 & 0 & 0 & 0 & \frac{GJ_S}{L_S} & 0 & 0 \\
0 & 0 & \frac{-6EI_Y}{L_S^2} & 0 & \frac{2EI_Y}{L_S} & 0 & 0 & 0 & \frac{6EI_Y}{L_S^2} & 0 & \frac{4EI_Y}{L_S} & 0 \\
0 & \frac{6EI_Z}{L_S^2} & 0 & 0 & 0 & \frac{2EI_Z}{L_S} & 0 & \frac{-6EI_Z}{L_S^2} & 0 & 0 & 0 & \frac{4EI_Z}{L_S}
\end{bmatrix}, u = \begin{bmatrix} u_{X_1}^S \\ u_{Y_1}^S \\ u_{Z_1}^S \\ \theta_{X_1}^S \\ \theta_{Z_1}^S \\ \theta_{Y_1}^S \\ u_{X_2}^S \\ u_{Y_2}^S \\ u_{Z_2}^S \\ \theta_{X_2}^S \\ \theta_{Z_2}^S \\ \theta_{Y_2}^S \end{bmatrix}$$ (41)

Using similar formulations, the mass and stiffness matrix for the first blade can be written as

$$M^{B'} =$$

$$\begin{bmatrix}
\frac{\rho A_B L_B}{3} & 0 & 0 & 0 & 0 & 0 & \frac{\rho A_B L_B}{6} & 0 & 0 & 0 & 0 & 0 \\
0 & \frac{156\rho A_B L_B}{420} & 0 & 0 & 0 & \frac{22\rho A_B L_B^2}{420} & 0 & \frac{54\rho A_B L_B}{420} & 0 & 0 & 0 & \frac{-13\rho A_B L_B^2}{420} \\
0 & 0 & \frac{156\rho A_B L_B}{420} & 0 & \frac{22\rho A_B L_B^2}{420} & 0 & 0 & 0 & \frac{54\rho A_B L_B}{420} & 0 & \frac{-13\rho A_B L_B^2}{420} & 0 \\
0 & 0 & 0 & \frac{\rho J_B L_B}{3} & 0 & 0 & 0 & 0 & 0 & \frac{\rho J_B L_B}{6} & 0 & 0 \\
0 & 0 & \frac{22\rho A_B L_B^2}{420} & 0 & \frac{4\rho A_B L_B^3}{420} & 0 & 0 & 0 & \frac{13\rho A_B L_B^2}{420} & 0 & \frac{-3\rho A_B L_B^3}{420} & 0 \\
0 & \frac{22\rho A_B L_B^2}{420} & 0 & 0 & 0 & \frac{4\rho A_B L_B^3}{420} & 0 & \frac{13\rho A_B L_B^2}{420} & 0 & 0 & 0 & \frac{-3\rho A_B L_B^3}{420} \\
\frac{\rho A_B L_B}{6} & 0 & 0 & 0 & 0 & 0 & \frac{\rho A_B L_B}{3} & 0 & 0 & 0 & 0 & 0 \\
0 & \frac{54\rho A_B L_B}{420} & 0 & 0 & 0 & \frac{13\rho A_B L_B^2}{420} & 0 & \frac{156\rho A_B L_B}{420} & 0 & 0 & 0 & \frac{-22\rho A_B L_B^2}{420} \\
0 & 0 & \frac{54\rho A_B L_B}{420} & 0 & \frac{13\rho A_B L_B^2}{420} & 0 & 0 & 0 & \frac{156\rho A_B L_B}{420} & 0 & \frac{-22\rho A_B L_B^2}{420} & 0 \\
0 & 0 & 0 & \frac{\rho J_B L_B}{6} & 0 & 0 & 0 & 0 & 0 & \frac{\rho J_B L_B}{3} & 0 & 0 \\
0 & 0 & \frac{-13\rho A_B L_B^2}{420} & 0 & \frac{-3\rho A_B L_B^3}{420} & 0 & 0 & 0 & \frac{-22\rho A_B L_B^2}{420} & 0 & \frac{4\rho A_B L_B^3}{420} & 0 \\
0 & \frac{-13\rho A_B L_B^2}{420} & 0 & 0 & 0 & \frac{-3\rho A_B L_B^3}{420} & 0 & \frac{-22\rho A_B L_B^2}{420} & 0 & 0 & 0 & \frac{4\rho A_B L_B^3}{420}
\end{bmatrix}$$

(42)

$$K^{B'} =$$

$$\begin{bmatrix}
\frac{EA_B'}{L_B} & 0 & 0 & 0 & 0 & 0 & \frac{-EA_B'}{L_B} & 0 & 0 & 0 & 0 & 0 \\
0 & \frac{12EI_Z'}{L_B^3} & 0 & 0 & 0 & \frac{6EI_Z'}{L_B^2} & 0 & \frac{-12EI_Z'}{L_B^3} & 0 & 0 & 0 & \frac{6EI_Z'}{L_B^2} \\
0 & 0 & \frac{12EI_Y'}{L_B^3} & 0 & \frac{-6EI_Y'}{L_B^2} & 0 & 0 & 0 & \frac{-12EI_Y'}{L_B^3} & 0 & \frac{-6EI_Y'}{L_B^2} & 0 \\
0 & 0 & 0 & \frac{GJ_X'}{L_B} & 0 & 0 & 0 & 0 & 0 & \frac{\rho L_B J_X'}{6} & 0 & 0 \\
0 & 0 & \frac{-6EI_Y'}{L_B^2} & 0 & \frac{4EI_Y'}{L_B} & 0 & 0 & 0 & \frac{6EI_Y'}{L_B^2} & 0 & \frac{2EI_Y'}{L_B} & 0 \\
0 & \frac{6EI_Z'}{L_B^2} & 0 & 0 & 0 & \frac{4EI_Z'}{L_B} & 0 & \frac{-6EI_Z'}{L_B^2} & 0 & 0 & 0 & \frac{2EI_Z'}{L_B} \\
\frac{-EA_B'}{L_B} & 0 & 0 & 0 & 0 & 0 & \frac{EA_B'}{L_B} & 0 & 0 & 0 & 0 & 0 \\
0 & \frac{-12EI_Z'}{L_B^3} & 0 & 0 & 0 & \frac{-6EI_Z'}{L_B^2} & 0 & \frac{12EI_Z'}{L_B^3} & 0 & 0 & 0 & \frac{-6EI_Z'}{L_B^2} \\
0 & 0 & \frac{-12EI_Y'}{L_B^3} & 0 & \frac{6EI_Y'}{L_B^2} & 0 & 0 & 0 & \frac{12EI_Y'}{L_B^3} & 0 & \frac{6EI_Y'}{L_B^2} & 0 \\
0 & 0 & 0 & \frac{\rho L_B J_X'}{6} & 0 & 0 & 0 & 0 & 0 & \frac{GJ_X'}{L_B} & 0 & 0 \\
0 & 0 & \frac{-6EI_Y'}{L_B^2} & 0 & \frac{2EI_Y'}{L_B} & 0 & 0 & 0 & \frac{6EI_Y'}{L_B^2} & 0 & \frac{4EI_Y'}{L_B} & 0 \\
0 & \frac{6EI_Z'}{L_B^2} & 0 & 0 & 0 & \frac{2EI_Z'}{L_B} & 0 & \frac{-6EI_Z'}{L_B^2} & 0 & 0 & 0 & \frac{4EI_Z'}{L_B}
\end{bmatrix}, u =$$

$$\left( u_{X_1}^B \quad u_{Y_1}^B \quad u_{Z_1}^B \quad \theta_{X_1}^B \quad \theta_{Y_1}^B \quad \theta_{Z_1}^B \quad u_{X_2}^B \quad u_{Y_2}^B \quad u_{Z_2}^B \quad \theta_{X_2}^B \quad \theta_{Y_2}^B \quad \theta_{Z_2}^B \right)^T \tag{43}$$

Where $A_B' = A_B \beta_X$, $J_X' = J_X \beta_{XR}$

The transformation matrix for the blade is written as:

$$\mathbf{T} = \begin{bmatrix} \cos(90) & \cos(90 + \theta^{(j)}) & \cos(\theta^{(j)}) \\ \cos(90) & \cos(\theta^{(j)}) & \cos(90 - \theta^{(j)}) \\ \cos(180) & \cos(90) & \cos(90) \end{bmatrix} \tag{44}$$

where $\theta^{(j)}$ is calculated as shown in Eq. (29). The full transformation matrix can then be constructed as

$$T^f = \begin{bmatrix} T & 0 & 0 & 0 \\ 0 & T & 0 & 0 \\ 0 & 0 & T & 0 \\ 0 & 0 & 0 & T \end{bmatrix} \tag{45}$$

The transformation of the blade matrices into the global system can be written as

$$\boldsymbol{M}^B = {T^f}^T \boldsymbol{M}^{B'} T^f, \boldsymbol{C}^B = {T^f}^T \boldsymbol{C}^{B'} T^f, \boldsymbol{K}^B = {T^f}^T \boldsymbol{K}^{B'} T^f u^{B_n}(t), \boldsymbol{f} = {T^f}^T \boldsymbol{f}^{B_n}(t) \tag{46}$$

where $\boldsymbol{M}^{B'}, \boldsymbol{C}^{B'}$ and $\boldsymbol{K}^{B'}$ are fully populated system matrices in the local coordinate system.

The calculation of the force vector has been calculated in the following section. The force along the X direction can be calculated by using the shape function and acceleration produced by centrifugal force. However, the shape function is defined only in the domain $\in [0, L_e^B]$ where $L_e^B$ is the length of an element of the blade, and the centrifugal force is defined in the domain of the complete blade length, which is $\in [0, L^B]$, we have shifted the coordinate system of the centrifugal force appropriately for each blade before integration, as shown in the following equation

$$F_j = \int_0^{L_e^B} \rho_a A_b N_j \left(x + (i-1)L_e^B + \frac{d^D}{2}\right)\omega^2 dx \tag{47}$$

for $j \in [1,2]$ where $N_1 = \left(1 - \frac{x}{L_e^B}\right)$, $N_2 = \left(1 + \frac{x}{L_e^B}\right)$ and $i \in [1,2,3 \ldots n]$ where n is the total number of elements in the blade. Using a similar approach along Y direction, we can define the forces as follows

$$F_j = \int_0^{L_e^B} \rho_a A_b N_j (-\bar{F}_D \cos \varepsilon_b - \bar{F}_L \sin \varepsilon_b) dx \tag{48}$$

$$F_j = \int_0^{L_e^B} \rho_a A_b N_j \left(\left(-(0.5\rho_a(\left((x + (i-1)L_e^B + \frac{d^D}{2})\omega\right)^2 + V_\infty^2) C_D S\right) \cos \varepsilon_b - (0.5\rho_a(\left((x + (i-1)L_e^B + \frac{d^D}{2})\omega\right)^2 + V_\infty^2) C_L S \sin \varepsilon_b))\right) dx \tag{49}$$

for $j \in [1,2]$ where $N_1 = \left(1 - 3\frac{x^2}{L_e^{B^2}} + 2\frac{x^3}{L_e^{B^3}}\right)$, $N_2 = \left(\frac{3x^2}{L_e^{B^2}} - \frac{2x^3}{L_e^{B^3}}\right)$ and $i \in [1,2,3 \ldots n]$ where n is the total number of elements in the blade. The forces along Y direction create a moment about the Z axis which can be defined as

$$M_j = \int_0^{L_e^B} \rho_a A_b N_j \left(\left(-(0.5\rho_a(\left((x + (i-1)L_e^B + \frac{d^D}{2})\omega\right)^2 + V_\infty^2) C_D S\right) \cos \varepsilon_b - (0.5\rho_a(\left((x + (i-1)L_e^B + \frac{d^D}{2})\omega\right)^2 + V_\infty^2) C_L S \sin \varepsilon_b))\right) dx \tag{50}$$

for $j \in [1,2]$ where $N_1 = \left(x + \frac{x^3}{L_e^{B^2}} - \frac{2x^2}{L_e^B}\right)$, $N_2 = \left(\frac{x^3}{L_e^{B^2}} - \frac{x^2}{L_e^B}\right)$ and $i \in [1,2,3 \ldots n]$ where n is the total number of elements in the blade. Along Z direction, we can define the forces as follows

$$F_j = \int_0^{L_e^B} \rho_a A_b N_j (-\bar{F}_D \sin \varepsilon_b + \bar{F}_L \cos \varepsilon_b) dx \tag{51}$$

$$F_j = \int_0^{L_e^B} \rho_a A_b N_j \left(-(0.5\rho_a(\left((x + (i-1)L_e^B + \frac{d^D}{2})\omega\right)^2 + V_\infty) C_D S) \sin \varepsilon_b - (0.5\rho_a S(\left((x + (i-1)L_e^B + \frac{d^D}{2})\omega\right)^2 + V_\infty^2) C_L S \cos \varepsilon_b)\right) dx \tag{52}$$

for $j \in [1,2]$ where $N_1 = \left(1 - 3\frac{x^2}{L_e^{B^2}} + 2\frac{x^3}{L_e^{B^3}}\right)$, $N_2 = \left(\frac{3x^2}{L_e^{B^2}} - \frac{2x^3}{L_e^{B^3}}\right)$ and $i \in [1,2,3 \ldots n]$ where n is the total number of elements in the blade and $d^D$ is the outer diameter of the disk. Now the whole force vector can be populated using the terms as calculated above. The forces along Z direction create a moment about Y-axis which can be defined as follows

$$M_j = \int_0^{L_e^B} \rho_a A_b N_j \left(-(0.5\rho_a(\left((x + (i-1)L_e^B + \frac{d^D}{2})\omega\right)^2 + V_\infty^2)C_D S)\sin\varepsilon_b - (0.5\rho_a S(\left((x + (i-1)L_e^B + \frac{d^D}{2})\omega\right)^2 + V_\infty^2)C_L S \cos\varepsilon_b)\right) dx \quad (53)$$

for $j \in [1,2]$ where $N_1 = \left(x + \frac{x^3}{L_e^{B^2}} - \frac{2x^2}{L_e^B}\right)$, $N_2 = \left(\frac{x^3}{L_e^{B^2}} - \frac{x^2}{L_e^B}\right)$ and $i \in [1,2,3 \ldots n]$ where n is the total number of elements in the blade. Now the whole force vector can be populated using the terms as calculated above.